\documentclass[12pt]{article}
\PassOptionsToPackage{hyphens}{url}
\usepackage[pagebackref=true,colorlinks,bookmarks=false]{hyperref}
\usepackage{amsmath,amssymb,amsthm,mathtools,amsrefs}
\usepackage{mathpazo} 
\usepackage{eulervm}
\usepackage{csquotes}
\usepackage{tikz}
\usetikzlibrary{arrows.meta}
\usetikzlibrary{decorations.pathmorphing,shapes,patterns}
\usetikzlibrary{calc}
\usepackage{adjustbox}
\usepackage{stmaryrd}
\usepackage{fancyhdr}
\usepackage{soul}
\usepackage{dsfont}
\usepackage[normalem]{ulem}
\usepackage{caption}
\usepackage{wrapfig}
\usepackage{pictexwd, dcpic}
\usepackage{float}
\usepackage{enumerate}
\usepackage[all]{xy} 
\hypersetup{
	colorlinks=true,
    linktoc=all,
    linkcolor=blue,  
    citecolor=blue,
    }

\usepackage{tocbasic}
\DeclareTOCStyleEntry[
  beforeskip=.2em plus 1pt,
  pagenumberformat=\textbf
]{tocline}{section}

\makeatletter
\newcommand{\altqedhere}{%
  \ifmeasuring@\else\sbox0{\popQED}\fi
  \tag*{\qedsymbol}%
}
\makeatother
    
\usepackage[framemethod=TikZ]{mdframed}
\usepackage{titlesec}

\newmdtheoremenv[%
   outerlinewidth=2,
   roundcorner=10pt,
   leftmargin=0,
   rightmargin=0,
   innertopmargin=4pt,
   backgroundcolor=green!05,
   outerlinecolor=blue!30,
   ntheorem=true,
   ]{fancydefn}{Definition}

\newmdtheoremenv[%
   outerlinewidth=2,
   roundcorner=10pt,
   leftmargin=0,
   rightmargin=0,
   innertopmargin=-4pt,
   backgroundcolor=blue!05,
   outerlinecolor=violet!30,
   ntheorem=true,
   ]{fancyex}{Example}
   
\newmdtheoremenv[%
   outerlinewidth=2,
   roundcorner=10pt,
   leftmargin=0,
   rightmargin=0,
   innertopmargin=-4pt,
   backgroundcolor=red!05,
   outerlinecolor=red!30,
   ntheorem=true,
   ]{fancyprop}{Proposition}

   \newmdtheoremenv[%
   outerlinewidth=2,
   roundcorner=10pt,
   leftmargin=0,
   rightmargin=0,
   innertopmargin=-4pt,
   backgroundcolor=violet!05,
   outerlinecolor=red!30,
   ntheorem=true,
   ]{fancythm}{Theorem}

\newenvironment{theo}[2][]{%
\begin{theorem}%
[#1]%
\label{#2}}%
{\end{theorem}}
   
\newenvironment{lem}[2][]{%
\begin{lemma}%
[#1]%
\label{#2}}%
{\end{lemma}}
\newenvironment{rmk}[2][]{%
\begin{remark}%
[#1]%
\label{#2}}%
{\end{remark}}
\newenvironment{phys}[2][]{%
\begin{physics}%
[#1]%
\label{#2}}%
{\end{physics}}
\newenvironment{prop}[2][]{%
\begin{proposition}%
[#1]%
\label{#2}}%
{\end{proposition}}
\newenvironment{cor}[2][]{%
\begin{corollary}%
[#1]%
\label{#2}}%
{\end{corollary}}
\newenvironment{defn}[2][]{%
\begin{definition}%
[#1]%
\label{#2}}%
{\end{definition}}
\newenvironment{notat}[2][]{%
\begin{notation}%
[#1]%
\label{#2}}%
{\end{notation}}
\newenvironment{exa}[2][]{%
\begin{example}%
[#1]%
\label{#2}}%
{\end{example}}
{\end{question}}
\numberwithin{equation}{section}

\makeatletter
\renewcommand*\env@matrix[1][\arraystretch]{%
  \edef\arraystretch{#1}%
  \hskip -\arraycolsep
  \let\@ifnextchar\new@ifnextchar
  \array{*\c@MaxMatrixCols c}}
\makeatother

\theoremstyle{plain}
\newtheorem{theorem}[equation]{Theorem}
\newtheorem{lemma}[equation]{Lemma}
\newtheorem{proposition}[equation]{Proposition}
\newtheorem{corollary}[equation]{Corollary}
\theoremstyle{definition}
\newtheorem{definition}[equation]{Definition}
\newtheorem{construction}[equation]{Construction}

\newtheorem{question}[equation]{Question}
\newtheorem{problem}[equation]{Problem}
\newtheorem{example}[equation]{Example}
\newtheorem{exercise}[equation]{Exercise}
\newtheorem*{answer}{Answer}
\newtheorem*{solution}{Solution}
\newtheorem{remark}[equation]{Remark}
\newtheorem{physics}[equation]{Physics}
\newtheorem{notation}[equation]{Notation}

\renewcommand\qedsymbol{$\blacksquare$}

\newcommand\define[1]{\emph{\textbf{#1}}}

\let\a=\alpha \let\b=\beta \let\g=\gamma \let\de=\delta 
\let\z=\zeta    
\let\l=\lambda 
\let\s=\sigma \let\t=\tau    
\let\w=\omega       \let\D=\Delta  
     
\let\C=\Chi 
\def\vf{\varphi}

\newcommand{\be}{\begin{equation}}
\newcommand{\ee}{\end{equation}}
\def\ba{\begin{align}} 
\def\ea{\end{align}}
\newcommand{\bea}{\begin{eqnarray}}
\newcommand{\eea}{\end{eqnarray}}
\newcommand{\bx}{\begin{example}}
\newcommand{\ex}{\end{example}}
\newcommand{\bex}{\begin{exercise}}
\newcommand{\eex}{\end{exercise}}
\newcommand{\ban}{\begin{answer}}
\newcommand{\ean}{\end{answer}}
\newcommand{\bt}{\begin{theorem}}
\newcommand{\et}{\end{theorem}}
\newcommand{\bc}{\begin{corollary}}
\newcommand{\ec}{\end{corollary}}
\newcommand{\blem}{\begin{lemma}}
\newcommand{\elem}{\end{lemma}}
\newcommand{\bp}{\begin{problem}}
\newcommand{\ep}{\end{problem}}
\newcommand{\bn}{\begin{proposition}}
\newcommand{\en}{\end{proposition}}
\newcommand{\bd}{\begin{definition}}
\newcommand{\ed}{\end{definition}}
\newcommand{\bcon}{\begin{construction}}
\newcommand{\econ}{\end{construction}}
\newcommand{\bq}{\begin{question}}
\newcommand{\eq}{\end{question}}
\newcommand{\bprf}{\begin{proof}}
\newcommand{\eprf}{\end{proof}}
\newcommand{\br}{\begin{remark}}
\newcommand{\er}{\end{remark}}
\newcommand{\bs}{\begin{solution}}
\newcommand{\es}{\end{solution}}
\newcommand{\beqs}{\begin{eqnarray}}
\newcommand{\eeqs}{\end{eqnarray}}

\newcommand{\id}{\mathrm{id}}

\newcommand{\mC}{\mathcal{C}}
\newcommand{\mM}{\mathcal{M}}
\newcommand{\mD}{\mathcal{D}}

\newcommand{\tr}{{\rm tr} }

\def\R{{{\mathbb R}}}
\def\C{{{\mathbb C}}}

\def\N{{{\mathbb N}}}
\def\Z{{{\mathbb Z}}}

\def\B{{{\mathbb B}}}

\def\Hi{{{\mathcal{H}}}}

\newcommand{\fdCAlg}{\mathbf{fdC\text{*-}Alg}}

\def\mA{{{\mathcal{A}}}}
\def\mS{{{\mathcal{S}}}}
\def\mB{{{\mathcal{B}}}}

\newcommand{\op}{\mathrm{op}}

\newcommand{\FinSet}{\mathbf{FinSet}}

\newcommand{\FinProb}{\mathbf{FinProb}}

\newcommand{\NCFP}{\mathbf{NCFinProb}}

\newcommand{\stoch}{\;\xy0;/r.25pc/:(-3,0)*{}="1";(3,0)*{}="2";{\ar@{~>}"1";"2"|(1.06){\hole}};\endxy\!}
\newcounter{sarrow}
\newcommand\xstoch[1]{%
\stepcounter{sarrow}%
\mathrel{\begin{tikzpicture}[baseline= {( $ (current bounding box.south) + (0,-0.1ex) $ )}]
\node[inner sep=.5ex] (\thesarrow) {\;$\scriptstyle #1$\;};
\path[draw,{<[scale=1.5,width=3,length=2]}-,decorate,
  decoration={snake,amplitude=0.3mm,segment length=2.1mm,pre=lineto,pre length=1pt}] 
    (\thesarrow.south east) -- (\thesarrow.south west);
\end{tikzpicture}}%
}

\newcommand{\bigboxplus}{
  \mathop{
    \vphantom{\bigoplus} 
    \mathchoice
      {\vcenter{\hbox{\resizebox{\widthof{$\displaystyle\bigoplus$}}{!}{$\boxplus$}}}}
      {\vcenter{\hbox{\resizebox{\widthof{$\bigoplus$}}{!}{$\boxplus$}}}}
      {\vcenter{\hbox{\resizebox{\widthof{$\scriptstyle\oplus$}}{!}{$\boxplus$}}}}
      {\vcenter{\hbox{\resizebox{\widthof{$\scriptscriptstyle\oplus$}}{!}{$\boxplus$}}}}
  }\displaylimits 
}

\newcommand{\ben}{\renewcommand{\theenumi}{\alph{enumi}} 
\renewcommand{\labelenumi}{(\theenumi)}\begin{enumerate}}
\newcommand{\een}{\end{enumerate}}

\newcommand\blfootnote[1]{%
  \begingroup
  \renewcommand\thefootnote{}\footnote{#1}%
  \addtocounter{footnote}{-1}%
  \endgroup
}

\title{A functorial characterization of von~Neumann entropy}
\author{Arthur~J.\ Parzygnat}
\date{\today}

\newcommand{\Addresses}{{
  \bigskip
  \footnotesize

  A.~Parzygnat, \textsc{Institut des Hautes \'Etudes Scientifiques, 35 Route de Chartres 91440, Bures-sur-Yvette, France}\par\nopagebreak
  \textit{E-mail address}, A.~Parzygnat: \texttt{parzygnat@ihes.fr}
}}

\begin{document}
\emergencystretch 2em
\maketitle 

\cfoot{}
\thispagestyle{empty}

\vspace{-9mm}

\begin{abstract} 
Using convex Grothendieck fibrations, we characterize the von~Neumann entropy as a functor from finite-dimensional non-commutative probability spaces and state-preserving $*$-homomorphisms to real numbers.
Our axioms reproduce those of Baez, Fritz, and Leinster characterizing the Shannon entropy difference. The existence of disintegrations for classical probability spaces plays a crucial role in our characterization. 
\blfootnote{\emph{2020 Mathematics Subject Classification.}  
18D30, 
81P17 (Primary); 
18C40, 
46L53, 
81R15, 
94A17 (Secondary). 
}
\blfootnote{
\emph{Key words and phrases.} 
Convex category, 
disintegration, 
Grothendieck fibration,
Landauer's principle, 
optimal hypothesis, 
quantum entropy
}
\end{abstract}

\tableofcontents

\section[Introduction and outline]{Introduction and outline}
\label{sec:intro}

In 2011, Baez, Fritz, and Leinster (BFL) characterized the Shannon entropy (difference) of finite probability distributions as the only non-vanishing continuous affine functor $\FinProb\to\B\R_{\ge0}$ from finite probability spaces to non-negative numbers up to an overall non-negative constant~\cite{BFL}. Here, $\FinProb$ is the category of finite sets equipped with probability measures as objects and probability-preserving functions as morphisms. The codomain category, $\B\R_{\ge0}$, is the category consisting of a single object and whose morphisms from that object to itself are all non-negative real numbers equipped with addition as the composition. 

A natural follow-up question is whether the von~Neumann (or finite-dimensional Segal) entropy can be characterized in a similar manner by replacing $\FinProb$ with $\NCFP$, the category of finite quantum (i.e.\ non-commutative) probability spaces, consisting of unital finite-dimensional $C^*$-algebras equipped with states as objects and state-preserving unital $*$-homomorphisms as morphisms. Physically, such objects correspond to hybrid classical/quantum systems and the morphisms describe deterministic dynamics, which includes tracing out subsystems. Although this question was partially explored by Baez and Fritz~\cite{BaezIII}, a suitably similar set of axioms was never obtained. The present manuscript accomplishes this task. 

There are two difficulties with extending BFL's result to the quantum setting. The first issue is that the difference of von~Neumann entropies need not have a fixed sign. There are state-preserving unital $*$-homomorphisms that \emph{increase} the entropy as well as \emph{decrease} the entropy. The sign of the entropy difference is closely related to the fact that Landauer's principle holds for classical systems~\cite{Lan61}, but could fail for quantum systems~\cites{del11,ReWo14}. The root of the increase stems from the uncertainty principle and entanglement. 

Using our axioms, we show that the existence of disintegrations~\cite{PaRu19} (called optimal hypotheses in~\cite{BaFr14}) implies the non-negativity of the entropy difference. Since disintegrations always exist for finite-dimensional classical systems, this proves one of the key assumptions of BFL in their functorial characterization of the Shannon entropy~\cite{BFL}.  

The second difficulty when attempting to extend BFL's work to quantum systems is that the objects of $\NCFP$ are not convex generated by any single object in that category. Note that this occurs for $\FinProb$, where an arbitrary probability space $(X,p)$, with $X$ a finite set and $p$ a probability measure on $X$, can be decomposed into a convex sum as
$
(X,p)\cong\bigoplus_{x\in X}p_{x}\mathbf{1},
$
where $\mathbf{1}$ is the (essentially) unique probability space consisting of a single element and $p_{x}$ is the probability of $x\in X$. In $\NCFP$, a quantum probability space such as $(\mathcal{M}_{m},\omega)$ cannot be expressed as a convex combination of lower-dimensional probability spaces. Here, $m\in\N$, $\mathcal{M}_{m}$ is the $C^*$-algebra of $m\times m$ matrices, and $\omega$ is a state on $\mathcal{M}_{m}$.

In this manuscript, we simultaneously address both these issues and provide a functorial characterization of the von~Neumann entropy. This is done by introducing Grothendieck fibrations of convex categories and fibred affine functors. The category $\NCFP$ forms a fibration over $\fdCAlg$, the category of finite-dimensional unital $C^*$-algebras and unital $*$-homomorphisms, by sending each quantum probability space $(\mA,\omega)$ to the underlying $C^*$-algebra $\mA$. The von~Neumann entropy (difference) provides a functor 
\be
\label{eq:entropydifferencefunctor}
\xy0;/r.20pc/:
(-15,7.5)*+{\NCFP}="1";
(15,7.5)*+{\B\R}="2";
(-15,-7.5)*+{\fdCAlg}="3";
(15,-7.5)*+{\uline{\mathbf{1}}}="4";
{\ar"1";"2"^(0.65){H}};
{\ar"1";"3"};
{\ar"2";"4"};
{\ar"3";"4"};
\endxy
\;\;,
\ee
where $\uline{\mathbf{1}}$ is the category consisting of a single object and just the identity morphism, $\B\R$ is the one-object category whose morphisms consist of all real numbers with composition rule given by addition, and the left vertical arrow is the fibration just mentioned. 

The fibres of the left and right fibrations in~(\ref{eq:entropydifferencefunctor}) are convex categories. Over each $C^*$-algebra $\mA$ on the left, one has the convex set of states $\mathcal{S}(\mA)$ on $\mA$, which is viewed as a discrete convex category. A morphism $f:\mB\rightarrow\mA$ of $C^*$-algebras gets lifted to the morphism $\mathcal{S}(f):\mathcal{S}(\mA)\to\mathcal{S}(\mB)$ that acts as the pullback of states, sending $\omega$ to $\omega\circ f$. On the right, $\B\R$ is also a convex category, with convex combinations of real numbers as the convex operation. 

This entropy difference functor sends a state $\omega\in\mS(\mA)$ together with a morphism $f:\mB\rightarrow\mA$ to a real number $H_{f}(\omega)$. Given another state $\xi\in\mS(\mA)$ and a number $\l\in[0,1]$, one obtains the inequality 
\be
\label{eq:Holevochangeinequality}
H_{f}\big(\l\omega+(1-\l)\xi\big)\ge \l H_{f}(\omega)+(1-\l)H_{f}(\xi),
\ee
which is of fundamental importance in quantum information theory. 
The non-negativity of the quantity 
\be
\chi_{f}(\l;\omega,\xi):=H_{f}\big(\l\omega+(1-\l)\xi\big)-\l H_{f}(\omega)-(1-\l)H_{f}(\xi)
\ee
is related to the \emph{monotonicity of entropy under partial trace}, which is known to be equivalent to \emph{strong subadditivity}~\cite{Ve13}. 
A special case of this inequality, when $f:=!_{\mA}:\C\to\mA$ is the unique unital $*$-homomorphism into $\mA$, leads to the fact that mixing always increases entropy. It is actually only this weaker property that will play a role in our current characterization. 

For more general algebras, if $\omega$ and $\xi$ have orthogonal supports, and if $f:\mB\to\mA$ preserves this orthogonality, then equality in~(\ref{eq:Holevochangeinequality}) is obtained. 
This condition, which we call \emph{orthogonal affinity}, is what replaces the affine assumption of entropy difference made by BFL. 
However, orthogonal affinity and~(\ref{eq:Holevochangeinequality}) are not enough to guarantee that $H_{\mA}(\w):=H_{!_{\mA}}(\w)$ vanishes on pure states $\omega$. If one imposes this additional assumption, one can show that it is no longer necessary to assume $\chi_{f}(\l;\omega,\xi)\ge0$ for all inputs. Instead, one can demand the simpler assumption that $H_{\mA}(\w)\ge0$ for all states $\w$. In other words, one can replace BFL's non-negativity assumption for classical entropy \emph{difference} with the assumption that $H_{\mA}(\w)\ge0$ for all states $\w$ on $C^*$-algebras $\mA$, with equality for pure states. The relationships between these assumptions will be made precise in the body of the present manuscript.
Our main theorem can then be phrased as follows.

\begin{theo}[A functorial characterization of quantum entropy (Theorem~\ref{thm:abstractentropy} in body)]{thm:main}
Let $H:\NCFP\to\B\R$ be a continuous and orthogonally affine fibred functor, 
as in~(\ref{eq:entropydifferencefunctor}), 
for which $H_{\mA}(\w)\ge0$ for all states $\w\in\mS(\mA)$, with equality on all pure states, for all finite-dimensional $C^*$-algebras $\mA$.
Then there exists a constant $c\ge0$ such that 
\[
H_{f}(\omega)=c\Big(S(\omega)-S(\omega\circ f)\Big)
\]
for all $*$-homomorphisms $\mB\xrightarrow{f}\mA$ of finite-dimensional $C^*$-algebras and states $\omega\in\mS(\mA)$. 
\end{theo}

In this theorem, $S(\omega)$ is the von~Neumann entropy of $\omega$, which is given by 
$
S(\omega)=-\tr(\rho\log\rho)
$
in the special case when $\omega=\tr(\rho\;\cdot\;)$ is a state on $\mathcal{M}_{m}$ represented by a unique density matrix $\rho$, with $\tr$ the (un-normalized) trace and $\cdot$ signifying the input of the function, i.e.\ $\mM_{m}\ni A\mapsto\tr(\rho A)$. More generally, when $\mA:=\bigoplus_{x\in X}\mathcal{M}_{m_{x}}$, a state $\omega$ on $\mA$ can be described by a collection of states $\omega_{x}\in\mS(\mathcal{M}_{m_{x}})$ and a probability measure $p$ on $X$ such that $\omega(A_{x})=p_{x}\omega_{x}(A_{x})$ for $A_{x}\in \mathcal{M}_{m_{x}}$. In this case, the entropy of $\omega$ is 
\be
S(\omega)=-\sum_{x\in X}p_{x}\log(p_{x})-\sum_{x\in X}p_{x}\tr(\rho_{x}\log\rho_{x}).
\ee
Since all finite-dimensional unital $C^*$-algebras are of this form (up to isomorphism), this specifies the functor $H$ everywhere, since entropy is invariant under isomorphism. 


The present manuscript is broken up as follows. We begin by reviewing states, mutual orthogonality, and entropy in Section~\ref{sec:states}. In particular, we provide translations between some operator-algebraic and physical concepts. 
Section~\ref{sec:fibrations} introduces fiberwise convex structures, fibered functors, and continuity of fibered functors.
Section~\ref{sec:classifyentropy} contains our main result and several others of potential interest. In particular, we prove that our axioms imply the non-negativity of $H_{f}(\w)$ for commutative $C^*$-algebras by using the fact that disintegrations exist for morphisms of commutative probability spaces. More generally, we prove that if a disintegration of $(f,\w)$ exists for an arbitrary quantum probability space $(\mA,\omega)$, then $H_{f}(\w)\ge0$. We also include a brief historical account of axiomatizations of the von~Neumann entropy and how our characterization compares with some of them. 


\section[States on finite-dimensional $C^*$-algebras]{States on finite-dimensional $C^*$-algebras}
\label{sec:states}

In this section, we set up notation and compile several standard facts that will be used throughout. All $C^*$-algebras will be unital and finite-dimensional and all $*$-homomorphisms will be unital unless stated otherwise. We will work in the Heisenberg picture, as will be explained in Example~\ref{ex:partialtrace}. Since all of our $C^*$-algebras will be finite-dimensional, they will always be $*$-isomorphic to direct sums of matrix algebras, so that most of our analysis will involve only linear algebra. An especially suitable reference including more than enough background is Farenick's linear algebra text~\cite{Fa01} (see Theorem~5.20 and Proposition~5.26 in~\cite{Fa01} for the statement regarding all finite-dimensional $C^*$-algebras).

\begin{defn}[Basic definitions]{defn:states}
Given a $C^*$-algebra $\mA,$ an element $a\in\mA$ is \define{positive} iff there exists an $x\in\mA$ such that $a=x^*x.$
The set of positive elements in $\mA$ 
is denoted by $\mA^{+}.$
An element $a\in\mA$ is \define{self-adjoint} iff $a^*=a$. 
An element $p\in\mA$ is a \define{projection} iff $p^*p=p$. 
The \define{orthogonal complement} of a projection $p\in\mA$ is the element $p^{\perp}:=1_{\mA}-p$ (and is also a projection). 
Positivity defines a partial order on self-adjoint elements and one writes $a\ge a'$ or $a'\le a$ iff $a-a'\in\mA^{+}$.
Given another $C^*$-algebra $\mB,$ a \define{positive map}%
\footnote{Motivated by stochastic Gelfand--duality~\cites{Pa17,FuJa15}, $*$-homomorphisms are always drawn with straight arrows $\to$, while linear maps between algebras are drawn with squiggly arrows $\stoch$, in order to distinguish between deterministic maps and stochastic maps.}
$\mB\xstoch{\vf}\mA$ is a linear map such that 
$\vf(\mB^{+})\subseteq\mA^{+}.$
A \define{weight} on a $C^*$-algebra $\mA$ is a positive map $\mA\xstoch{\omega}\C$. 
A weight is called a \define{state} iff it is unital. The set of states on a $C^*$-algebra $\mA$ are denoted by $\mathcal{S}(\mA)$. 

A \define{non-commutative/quantum probability space} %
 is a pair $(\mA,\w)$ consisting of a $C^*$-algebra together with a state $\w\in\mS(\mA)$. A \define{state-preserving} map (a $*$-homomorphism or a positive map) from one non-commutative probability space $(\mB,\xi)$ to another $(\mA,\w)$ is a map $\mB\xstoch{f}\mA$ such that $\xi=\w\circ f$. 
A state $\w\in\mS(\mA)$ is \define{pure} iff it cannot be expressed as a non-trivial convex combination of some pair of distinct states. 
For the $C^*$-algebra of $m\times m$ matrices $\mathcal{M}_{m}$, which is referred to as a \define{matrix algebra}, the involution is the conjugate transpose and is denoted by $\dag$ instead of $*$. If $m=1$, then $\mM_{1}\cong\C$ and  $\overline{z}$ is used to denote the complex conjugate of $z\in\C$. 
\end{defn}

\begin{exa}[Density matrices, states, and expectation values]{ex:expectationvalues}
Self-adjointness and positive semidefiniteness of an $m\times m$ matrix coincides with the $C^*$-algebraic definition of positivity on $\mM_{m}$. 
Every state $\omega$ on $\mM_{m}$ can be expressed as $\omega=\tr(\rho\;\cdot\;)$ for some unique \define{density matrix} $\rho\in\mM_{m}$, which is a positive matrix such that $\tr(\rho)=1$. Here, and everywhere else in this manuscript, $\tr$ denotes the un-normalized trace. 

When $\mA:=\bigoplus_{x\in X}\mathcal{M}_{m_{x}}$, with $X$ a finite set and $m_{x}\in\N$, a state $\omega$ on $\mA$ can be described by a collection of states $\omega_{x}\in\mS(\mathcal{M}_{m_{x}})$ and a probability measure $p$ on $X$ such that $\omega(A_{x})=p_{x}\omega_{x}(A_{x})$ for $A_{x}\in \mathcal{M}_{m_{x}}$~\cite[Lemma~5.27]{PaRu19}. Here, and elsewhere in the manuscript, $p_{x}$ is used to denote the probability of $x$ with respect to $p$. 
Since each state $\omega_{x}$ corresponds to a density matrix $\rho_{x}\in\mM_{m_{x}},$ $\omega$ can equivalently be expressed as $\omega(A_{x})=p_{x}\tr(\rho_{x}A_{x})$ for $A_{x}\in\mathcal{M}_{m_{x}}$. We will also use all of the following notations
\[
\w\equiv\sum_{x\in X}p_{x}\w_{x}\equiv\sum_{x\in X}p_{x}\tr(\rho_{x}\;\cdot\;)
\]
to indicate the same state. 
In this way, states encode the data of families of expectation values. Since every $C^*$-algebra $\mA$ is isomorphic to a finite direct sum of matrix algebras, this is a full description of states on $C^*$-algebras. 

The usefulness of using $C^*$-algebras as opposed to just matrix algebras is to allow for a combination of classical and quantum setups, such as measurement. Furthermore, direct sums of matrix algebras are used in describing superselection sectors~\cites{Wi18,Pe01}, while ensembles, preparations, instruments, etc.\ are all naturally described by positive maps between certain $C^*$-algebras that are not just matrix algebras~\cite[Section~4]{PaRuBayes}. 
\end{exa}

\begin{lem}[The support of a weight]{lem:support}
Associated to every weight $\omega$ on a $C^*$-algebra $\mA$ is a projection $P_{\omega}\in\mA$ satisfying %
\[
\omega(P_{\omega}A)=\omega(AP_{\w})=\w(P_{\w}AP_{\w})=\w(A)\qquad\forall\;A\in\mA
\]
and such that $P_{\w}\le Q$ for every other projection $Q$ satisfying this condition (with $Q$ replacing $P_{\w}$). 
\end{lem}

\begin{defn}[Supports and mutually orthogonal weights]{defn:orthogonality}
The projection $P_{\w}$ in Lemma~\ref{lem:support} is called the \define{support} of $\w$. 
Two weights $\omega,\xi$ on a finite-dimensional $C^*$-algebra $\mA$ are \define{mutually orthogonal}, written $\omega\perp\xi$, iff any of the following equivalent conditions hold.%
\footnote{For the thermodynamic meaning of mutual orthogonality of states, see~\cite[Section~2]{Pe01}.}
\begin{enumerate}
\itemsep0pt
\item
If for any weight $\chi$ on $\mA$ such that $\chi\le\omega$ and $\chi\le\xi$, then $\chi=0$. 
\item
$P_{\w}P_{\xi}=0$ (which implies $P_{\w}P_{\xi}=P_{\xi}P_{\w}$).
\end{enumerate}
A $*$-homomorphism $\mB\xrightarrow{f}\mA$ \define{preserves the mutual orthogonality $\w\perp\xi$} iff $(\w\circ f)\perp(\xi\circ f)$. 
\end{defn}

\begin{lem}[The image of a support]{lem:imagesofsupports}
Let $\mB\xrightarrow{f}\mA$ be a $*$-homomorphism and let $\mA\xstoch{\w}\C$ be a state. Then 
$f(P_{\w\circ f}^{\perp})\le P_{\w}^{\perp}$ and $f(P_{\w\circ f})\ge P_{\w}$.
\end{lem}

\bprf
The first inequality follows from the fact that $f$ sends projections to projections and $f(\mathcal{N}_{\w\circ f})\subseteq\mathcal{N}_{\w}$ (see the proof of \cite[Proposition~3.2]{PaRu19}), where 
\be
\mathcal{N}_{\xi}:=\{A\in\mA\;:\;\xi(A^*A)=0\}
\ee
denotes the nullspace associated to a state $\xi$. 
The two inequalities are equivalent because
\be
\begin{split}
f(P_{\w\circ f})&=f(1_{\mB}-P_{\w\circ f}^{\perp})=f(1_{\mB})-f(P_{\w\circ f}^{\perp})\\
&=1_{\mA}-f(P_{\w\circ f}^{\perp})=f(P_{\w\circ f}^{\perp})^{\perp}\ge P_{\w}, 
\end{split}
\ee
where the last inequality used $f(P_{\w\circ f}^{\perp})\le P_{\w}^{\perp}$. 
A similar calculation shows the converse. 
\eprf

\begin{exa}[External convex sums for finite probability spaces]{ex:BFLexternal}
Let $X,X',Y,Y'$ be finite sets, let $p$ and $q$ be probability measures on $X$ and $Y$, respectively, and let $X\xrightarrow{\phi}X'$ and $Y\xrightarrow{\psi}Y'$ be two functions. 
Let $\l p\oplus(1-\l)q$ denote the probability measure on $X\amalg Y$ (the disjoint union) given by 
\[
(\l p\oplus(1-\l)q)_{z}:=\begin{cases}\l p_{z}&\mbox{ if } z\in X\\ (1-\l)q_{z}&\mbox{ if }z\in Y\end{cases}. 
\]
Set $\mA:=\C^{X}$ and $\mB:=\C^{Y}$ to be the $C^*$-algebras of functions on $X$ and $Y$, and similarly $\mA':=\C^{X'}$ and $\mB':=\C^{Y'}$. 
Let $\omega$ and $\xi$ be the states on $\mA$ and $\mB$ associated to $p$ and $q$, i.e.\ 
 $\w(A)=\sum_{x\in X}p_{x}A(x)$ for all $A\in\C^{X}$ (and similarly for $\xi$ and $q$).  
Let $\mA'\xrightarrow{f}\mA$ and $\mB'\xrightarrow{g}\mB$ be the $*$-homomorphisms associated to $\phi$ and $\psi$ via pullback. Namely, if $A'\in\C^{X'}$ is a function on $X'$, then $f(A'):=A'\circ\phi$. The disjoint union function $X\amalg Y\xrightarrow{\phi\amalg\psi}X'\amalg Y'$ corresponds to the direct sum $*$-homomorphism 
\[
\C^{X'\amalg Y'}\cong\mA'\oplus\mB'\xrightarrow{f\oplus g}\mA\oplus\mB\cong\C^{X\amalg Y}.
\]
Let $\widetilde{\omega}$ and $\widetilde{\xi}$ denote the states on $\mA\oplus\mB$ given by $\widetilde{\omega}(A\oplus B):=\omega(A)$ and $\widetilde{\xi}(A\oplus B):=\xi(B)$ for all $A\in\mA$ and $B\in\mB$. From these definitions, the state on $\mA\oplus\mB$ associated to $\l p\oplus(1-\l)q$ is $\l\widetilde{\w}+(1-\l)\widetilde{\xi}$. Furthermore, $\widetilde{\omega}\perp\widetilde{\xi}$ holds and $f\oplus g$ preserves  $\widetilde{\omega}\perp\widetilde{\xi}$. This construction of convex sums is one of the main ingredients in BFL's characterization of entropy~\cite{BFL}. 
\end{exa}

\begin{notat}[Internal direct sum]{not:internalsum}
Let $m\in\N$, $Y$ a finite set, $\{n_{y}\}_{y\in Y}$ a collection of natural numbers satisfying $m=\sum_{y\in Y}n_{y}$, and $\{B_{y}\in\mM_{n_{y}}\}_{y\in Y}$ a collection of matrices. Given an ordering of the elements of $Y$, set 
\[
\bigboxplus_{y\in Y}B_{y}:=\begin{bmatrix}
B_{1}&&0\\
&\ddots&\\
0&&B_{|Y|}
\end{bmatrix}
\equiv
\mathrm{diag}(B_{1},\dots,B_{|Y|})\in\mM_{m}
.
\]
This notation will be frequently used, sometimes without explicitly stating that an order has been chosen.%
\footnote{This is not to be confused with the (external) direct sum $\bigoplus_{y\in Y}B_{y}\in\bigoplus_{y\in Y}\mM_{n_{y}},$ which does not use an ordering on $Y$ and, more importantly, is an element of a different (non-isomorphic) algebra.}
\end{notat}

\begin{exa}[The partial trace]{ex:partialtrace}
Working with unital $*$-homomorphisms between $C^*$-algebras corresponds to the \emph{Heisenberg picture} description of quantum mechanics, as opposed to the more commonly used \emph{Schr\"odinger picture} in the quantum information theory community. The relationship between the two goes roughly as follows. 

If $\mB=\mathcal{M}_{n}$, $\mA=\mM_{m}$, and $\mB\xrightarrow{f}\mA$ is a $*$-homomorphism, then there exists a $p\in\N$ such that $m=pn$ and a unitary $U\in\mM_{m}$ such that $f=\mathrm{Ad}_{U}\circ g$, where $\mathrm{Ad}_{U}(A):=UAU^{\dag}$ for all $A\in\mA$, and where $g$ is 
\[
\mB\ni B\mapsto g(B):=
\mathds{1}_{p}\otimes B
\]
(cf.\ \cite{Attal},~\cite[Lecture~10]{Werner17}). 
The adjoint, $g^*$, of $g$ with respect to the \define{Hilbert--Schmidt} or \define{Frobenius} inner product on the vector space of linear maps between $\mA$ and $\mB$ is given by
\[
\mA\cong\mM_{p}\otimes\mM_{n}\ni A\otimes B \mapsto g^*(A\otimes B)=\tr(A)B.
\]
It is often written as $\tr_{\mM_{p}}$ and is called the \define{partial trace} (see~\cite[Sections~2 and~4]{PaRu19} or~\cite[Section~2.4.3]{NiCh11} for more details). The adjoint of $f$ is $g^*\circ\mathrm{Ad}_{U^{\dag}}$. 
%
\end{exa}

\begin{lem}[The partial trace on direct sums]{lem:ptracedirectsums}
Let $\mB:=\bigoplus_{y\in Y}\mM_{n_{y}}\xrightarrow{f}\bigoplus_{x\in X}\mM_{m_{x}}=:\mA$ be a $*$-homomorphism and let $\w=\sum_{x\in X}p_{x}\tr(\rho_{x}\;\cdot\;)$ be a state on $\mA$ (cf.\ Example~\ref{ex:expectationvalues}). Then the following facts hold.  
\begin{enumerate}
\itemsep0pt
\item
There exists a collection $\{c_{xy}\}$ of non-negative numbers, with $c_{xy}$ called the \define{multiplicity} of the factor $\mathcal{M}_{n_{y}}$ inside $\mathcal{M}_{m_{x}}$ associated to $f$, such that $m_{x}=\sum_{y\in Y}c_{xy}n_{y}$ for all $x\in X$. 
\item
There exist unitaries $U_{x}\in\mM_{m_{x}}$ such that $f$ is of the form 
\[
\bigoplus_{y\in Y}\mM_{n_{y}}\ni \bigoplus_{y\in Y}B_{y}\xmapsto{f}\bigoplus_{x\in X}U_{x}\Bigg(\bigboxplus_{y\in Y}\mathrm{diag}(\overbrace{B_{y},\cdots,B_{y}}^{c_{yx}\text{ times}})\Bigg)U_{x}^{\dag}.
\]
\item
The pullback state $\xi:=\w\circ f$ can be expressed as 
\[
\xi=\sum_{y\in Y}q_{y}\tr(\s_{y}\;\cdot\;),\quad\text{ where }\quad
q_{y}\s_{y}=\sum_{x\in X}p_{x}f^*_{xy}(\rho_{x})\qquad\forall\;y\in Y
\]
and $f^*_{xy}$ denotes the (Hilbert--Schmidt) adjoint of $f_{xy}:\mM_{n_{y}}\to\mM_{m_{x}}$, which is the component of $f$ mapping between the factors as indicated.
\end{enumerate}
\end{lem}

\bprf
See~\cite[Sections~1.1.2 and 1.1.3]{Fi96}, \cite[Theorem~5.6]{Fa01}, and~\cite[Lemma~6.7]{PaRuBayes}.
\eprf


\begin{lem}[$*$-isomorphisms preserve mutual orthogonality]{lem:isopreserveperp}
Let $\mB\xrightarrow{f}\mA$ be a $*$-isomorphism and let $\w,\xi$ be any two states on $\mA$. Then 
$\w\perp\xi$ implies $(\w\circ f)\perp(\xi\circ f)$. 
Furthermore, 
$\z\in\mS(\mA)$ is pure if and only if $\z\circ f$ is pure. 
\end{lem}

\bprf
If $P_{\w}$ and $P_{\xi}$ are the supports of $\w$ and $\xi$, respectively, then the claim will follow if we prove $f^{-1}(P_{\w})$ and $f^{-1}(P_{\xi})$ are the supports of $\w\circ f$ and $\xi\circ f$, respectively, because 
\be
f^{-1}(P_{\w})f^{-1}(P_{\xi})=f^{-1}(P_{\w}P_{\xi})=f^{-1}(0)=0.
\ee 
It suffices to focus on $\w$. First, note that $f^{-1}(P_{\w})$ is a projection since $f^{-1}$ is a $*$-homomorphism. Furthermore, 
\be
(\w\circ f)\big(f^{-1}(P_{\w})B\big)
=\w\big(P_{\w}f(B)\big)
=\w\big(f(B)\big)=(\w\circ f)(B)
\ee
for all $B\in\mB$, 
which proves that $f^{-1}(P_{\w})$ satisfies the first condition of a support for $\w\circ f$ in Lemma~\ref{lem:support}. Suppose that $Q$ is another projection satisfying $(\w\circ f)(QB)=(\w\circ f)(B)$ for all $B\in\mB$. Then $f(Q)$ satisfies
\be
\w\big(f(Q)A\big)=(\w\circ f)\big(Qf^{-1}(A)\big)
=(\w\circ f)\big(f^{-1}(A)\big)
=\w(A)
\ee 
for all $A\in\mA$. 
Hence, since $P_{\w}$ is the minimal such projection, $P_{\w}\le f(Q)$. Since $*$-homomorphisms preserve the $\le$ order structure, $f^{-1}(P_{\w})\le Q$. 
\eprf

\begin{exa}[Channels that do not preserve orthogonality]{ex:channelsthatdontpreserveorthogonality}
There are many examples of $*$-homomorphisms $\mB\to\mA$ that do not always preserve mutual orthogonality. A simple example is $!_{\C^{2}}:\C\to\C^2$, where every pair of mutually orthogonal states gets pulled back to $1$. A non-classical example is the $*$-homomorphism $\mM_{2}\to\mM_{2}\otimes\mM_{2}$, sending $B$ to $B\otimes\mathds{1}_{2}$, and any two density matrices on $\C^{2}\otimes\C^{2}$ corresponding to any two orthogonal Bell states~\cite[Section~2.3]{NiCh11}. In either case, the pullback state is $\frac{1}{2}\tr$. 
\end{exa}
 
\begin{lem}[Overlapping states remain overlapping under evolution]{lem:overlappingstatesstayoverlapping}
Let $\mB\xrightarrow{f}\mA$ be $*$-homomorphism and let $\w,\xi$ be  two states on $\mA$ that are \emph{not} mutually orthogonal. Then $\w\circ f$ and $\xi\circ f$ are also not mutually orthogonal.  
\end{lem}

\bprf
Suppose, to the contrary, that $P_{\w\circ f}P_{\xi\circ f}=0$.
Then
\be
0=f(0)=f(P_{\w\circ f}P_{\xi\circ f})=f(P_{\w\circ f})f(P_{\xi\circ f}).
\ee
But, by Lemma~\ref{lem:imagesofsupports}, $f(P_{\w\circ f})\ge P_{\w}$ and $f(P_{\xi\circ f})\ge P_{\xi}$ so that their product cannot vanish by the assumption $P_{\w}P_{\xi}\ne0$. This is a contradiction.
\eprf

\begin{phys}[Evolving states with overlapping supports]{phys:evolvingoverlap}
The interpretation of Lemma~\ref{lem:overlappingstatesstayoverlapping} is that if two states have overlapping supports, then no quantum operation will ever completely separate them. In contrast, Lemma~\ref{lem:isopreserveperp} says that reversible dynamics (such as unitary evolution) cannot mix states. 
\end{phys}

Now that we have defined the objects and morphisms of interest, we can define entropy and its generalizations to matrix algebras and $C^*$-algebras. 

\begin{defn}[Shannon, von~Neumann, and Segal entropy]{defn:entropyquantum}
Let $\omega$ be a state on $\mA$ as in Example~\ref{ex:expectationvalues}. The \define{Segal entropy} of $\omega$ is the non-negative number
\[
S_{\mathrm{Se}}(\omega):=
S_{\mathrm{Sh}}(p)+\sum_{x\in X}p_{x}S_{\mathrm{vN}}(\rho_{x}),
\]
where
$
S_{\mathrm{Sh}}(p):=-\sum_{x\in X}p_{x}\log(p_{x})
$
is the \define{Shannon entropy} of a probability measure $p$ on $X$
and
$
S_{\mathrm{vN}}(\rho):=-\tr\big(\rho\log\rho\big)
$
is the \define{von~Neumann entropy} of a density matrix $\rho$ on $\C^{n}$. The convention $0\log0:=0$ is used. 

On occasion, 
the letter $S$ will exclusively be used to refer to any of these three entropies, using the input to distinguish which formula should be used. As such, \define{entropy} will refer to any of these three, while \define{quantum entropy} will refer to either $S_{\mathrm{Se}}$ or $S_{\mathrm{vN}}$.%
\footnote{The Segal entropy was actually defined much more generally for certain infinite-dimensional systems~\cite{Se60}. The Segal entropy also equals $S_{\mathrm{Se}}(\omega)=-\sum_{x\in X}\tr\big(p_{x}\rho_{x}\log(p_{x}\rho_{x})\big)$.} 
\end{defn}

We recall the following useful fact about the entropy of convex combinations. 

\begin{lem}[Concavity inequalities for entropy]{lem:derivationpropertyonorthogonal}
Let $\{\rho_{x}\}_{x\in X}$ be a collection of density matrices on a Hilbert space indexed by a finite set $X$. Then 
\[
\sum_{x\in X}p_{x}S_{\mathrm{vN}}(\rho_{x})\le S_{\mathrm{vN}}\left(\sum_{x\in X}p_{x}\rho_{x}\right)\le S_{\mathrm{Sh}}(p)+\sum_{x\in X}p_{x}S_{\mathrm{vN}}(\rho_{x})
\]
for any probability distribution $p$ on $X$. Furthermore, the second inequality becomes an equality if and only if $\rho_{x}\perp\rho_{x'}$ for all distinct $x,x'\in X$ such that $p_{x}\ne0$ and $p_{x'}\ne0$.
\end{lem}

\bprf
The first inequality is the concavity of the von~Neumann entropy. Proofs of these claims can be found in~\cite[Theorem~11.8~(4)]{NiCh11} as well~\cite[Corollary pg~247]{Li72} and \cite[Equation~(2.2)]{Li73}. 
\eprf


We now come to our main definition for the entropy change along a morphism.

\begin{defn}[The entropy change along a morphism]{defn:entropychange}
Let $\mB\xrightarrow{f}\mA$ be a $*$-homomorphism of $C^*$-algebras and let $\omega$ be a state on $\mA$. The \define{entropy change of $\omega$ along $f$} is the number
\[
S_{f}(\omega):=S_{\mathrm{Se}}(\omega)-S_{\mathrm{Se}}(\omega\circ f). 
\]
\end{defn}

The following lemma contains a crucial observation that distinguishes the entropy change along a morphism between commutative versus non-commutative $C^*$-algebras. 

\begin{lem}[The entropy change along certain morphisms]{lem:entropychangecommutative}
Recall the notation from Definition~\ref{defn:entropychange}. 
\begin{enumerate}
\itemsep0pt
\item
\label{item:fiso}
If $f$ is a $*$-isomorphism, then $S_{f}(\omega)=0$ for all states $\omega\in\mS(\mA)$. 
\item
If $\mA$ and $\mB$ are commutative $C^*$-algebras, then $S_{f}(\omega)\ge0$ for all states $\omega\in\mS(\mA)$ and $*$-homomorphisms $\mB\xrightarrow{f}\mA$. 
\item
If $\mA$ is not commutative and $f$ is not a $*$-isomorphism, then there exists a state $\omega\in\mS(\mA)$ such that $S_{f}(\omega)<0$.%
\footnote{If $\mB$ is not commutative, then a $*$-homomorphism $\mB\to\mA$ does not exist if $\mA$ is commutative. 
}
\end{enumerate}
\end{lem}

\bprf
{\color{white}{you found me!}}
\begin{enumerate}
\itemsep0pt
\item
Let $\mA,\mB,\omega,f$, and $\xi$ be as in Example~\ref{ex:expectationvalues}. Since $f$ is a $*$-isomorphism, there exists a bijection $X\xrightarrow{\phi}Y$ and a collection of unitaries $U_{x}\in\mM_{m_{x}}$ such that 
\be
m_{x}=n_{\phi(x)}\quad\text{ and }\quad
p_{x}U_{x}\rho_{x}U_{x}^{\dag}=q_{\phi(x)}\s_{\phi(x)}
\qquad\forall\;x\in X
\ee
by Lemma~\ref{lem:ptracedirectsums}. 
The claim $S_{f}(\w)=0$ then follows from the functional calculus and Definition~\ref{defn:entropyquantum}. 
\item
Since every commutative finite-dimensional $C^*$-algebra is isomorphic to functions on a finite set as described in Example~\ref{ex:BFLexternal}, the Segal entropy becomes the Shannon entropy. If $p$ and $q$ are the probability measures on $X$ and $Y$ corresponding to $\omega$ and $\w\circ f$, respectively, then 
\be
S_{f}(\omega)=S_{\mathrm{Se}}(\w)-S_{\mathrm{Se}}(\w\circ f)
=S_{\mathrm{Sh}}(p)-S_{\mathrm{Sh}}(q),
\ee
which is shown to be non-negative in~\cite{BFL} (see Proposition~\ref{prop:positivitycommutative} for a more general and abstract proof using disintegrations).
\item
If $\mA$ is not commutative, then it has some matrix algebra $\mM_{m}$ as a factor with $m>1$. Let $\rho$ be a rank 1 density matrix in $\mA$ with support in $\mM_{m}$ (so that $\rho$ is a pure state). Let $A$ be a self-adjoint $m\times m$ matrix that does not commute with $\rho$ (such a matrix necessarily exists because the center of $\mM_{m}$ consists of multiples of the identity). Let $\s(A)$ denote the spectrum of $A$. Let $\mB:=\C^{\s(A)}\xrightarrow{f}\mA$ send $e_{\l}$, the function on $\s(A)$ whose value at $\l$ is 1 and is 0 elsewhere, to $P_{\l}$ in $\mM_{m}$, the projection onto the $\l$-eigenspace. 
Then $\w\circ f$ is not a pure state, in the sense that the associated measure on $\s(A)$ is not a Dirac measure. Thus, the entropy change is $S_{f}(\omega)=S_{\mathrm{Se}}(\w)-S_{\mathrm{Se}}(\w\circ f)=0-S_{\mathrm{Se}}(\w\circ f)<0$. \qedhere
\end{enumerate}
\eprf

Item 2 in Lemma~\ref{lem:entropychangecommutative} was used as an axiom by BFL to characterize the entropy change in the classical setting. Since it fails when one includes non-commutative $C^*$-algebras, we will have to replace this axiom with one that more accurately reflects the properties of entropy in quantum mechanics. 

\begin{phys}[Negative conditional entropy]{phys:negcondentropy}
As another example illustrating the validity of item 3 in Lemma~\ref{lem:entropychangecommutative} using only matrix algebras, take $\w$ on $\mM_{2}\otimes\mM_{2}\cong\mM_{4}$ to be a Bell state and let $\mM_{2}\xrightarrow{f}\mM_{2}\otimes\mM_{2}$ be the inclusion into one of the factors. Then $S_{f}(\w)=-\log(2)$ (cf.\ Example~\ref{ex:channelsthatdontpreserveorthogonality}). 
More generally, set $\mA:=\mM_{m}$, $\mB:=\mM_{n}$, $\mA\xrightarrow{f}\mA\otimes\mB$ the standard inclusion, and $\w=\tr(\rho_{\mA\mB}\;\cdot\;)$, where $\rho_{\mA\mB}$ is a density matrix in $\mA\otimes\mB$ with marginals $\rho_{\mA}:=\tr_{\mB}(\rho_{\mA\mB})$ and $\rho_{\mB}:=\tr_{\mA}(\rho_{\mA\mB})$ (cf.\ Example~\ref{ex:partialtrace}). Then the entropy difference $S_{f}(\w)=S_{\mathrm{vN}}(\rho_{\mA\mB})-S_{\mathrm{vN}}(\rho_{\mA})$ is the \define{quantum conditional entropy}, which, if negative, necessarily implies that $\rho_{\mA\mB}$ is entangled (see near Equation~(21) in~\cite{HoHo94}). %
The example we chose in the proof of Lemma~\ref{lem:entropychangecommutative} is meant to illustrate that entanglement  
is not necessary for $S_{f}(\w)$ to be negative. 
\end{phys}

\begin{phys}[Information loss or gain and Landauer's principle]{phys:Landauer}
In~\cite{BFL}, BFL interpreted the non-negative entropy difference between \emph{commutative} algebras as information loss. Indeed, a state-preserving $*$-homomorphism between commutative probability spaces corresponds to a probability-preserving map of finite sets equipped with probabilities. Such a map may identify points in an irreversible manner (in the sense that a probability-preserving inverse need not exist). When two points get identified, the corresponding probabilities add (cf.\ Definition~\ref{defn:disintegration}) and there is a decrease in entropy. This is closely related to Landauer's principle~\cite{Lan61}, which states that erasure (information loss) entails the dissipation of energy (in the form of heat) into the environment. 

For non-commutative probability spaces, i.e.\ quantum systems, information and work can  be \emph{gained} in certain situations, violating Landauer's principle. The information can be later used for state merging protocols~\cites{HOW05,HOW07} or the corresponding energy can be used to do thermodynamic work~\cite{del11}. A precise reformulation of the principle has been recently stated and proved in the case of finite-dimensional matrix algebras~\cite{ReWo14}. 
\end{phys}

We now end this section with a summary of the categories that will be used throughout. 

\begin{notat}[Categories used in this work]{defn:categorynotation}
In all categories that follow, except the very last one, the composition rule will be function composition. 
\begin{enumerate}
\itemsep0pt
\item
$\FinSet$ is the category whose objects are finite sets and whose morphisms are functions. 
\item
$\FinProb$ is the category whose objects are \define{finite probability spaces}, which are pairs $(X,p)$, with $X$ a finite set and $p$ a probability measure on $X$. A morphism from $(X,p)$ to $(Y,q)$ is a \define{probability-preserving function}, i.e.\ a function $X\xrightarrow{\phi}Y$ such that $q_{y}=\sum_{x\in \phi^{-1}(\{y\})}p_{x}$ for all $y\in Y$, where $\phi^{-1}(\{y\}):=\{x\in X\,:\,\phi(x)=y\}$.
\item
$\fdCAlg$ is the category whose objects are (finite-dimensional unital) $C^*$-algebras and morphisms are (unital) $*$-homomorphisms. 
\item
\emergencystretch 3em
$\NCFP$ is the category whose objects are (finite-dimensional) non-commutative probability spaces and whose morphisms are state-preserving (unital) $*$-homomorphisms.
\item
$\B\R$ ($\B\R_{\ge0}$) is the category consisting of a single object and whose morphisms from that object to itself are all real numbers (non-negative real numers) equipped with addition as the composition rule.
\end{enumerate}
Finally, here are some additional categorical notations and terminologies that will be used.
Given two categories $\mC$ and $\mD$, let $\mC\times\mD$ denote their cartesian product. Let $\mC\times\mD\xrightarrow{\g}\mD\times\mC$ be the functor that swaps the two inputs. Let $\mC\xrightarrow{\D}\mC\times\mC$ be the diagonal functor sending an object $x$ to $(x,x)$ and similarly for morphisms. There are two projection functors, denoted by $\mC\times\mD\xrightarrow{\pi_{1}}\mC$ and $\mC\times\mD\xrightarrow{\pi_{2}}\mD$.
\end{notat}

\section[Fibrations and local convex structures]{Fibrations and local convex structures}
\label{sec:fibrations}

Fibrations provide a convenient setting to formulate the notion of entropy change as a functor. Non-commutative probability spaces form a discrete fibration over $C^*$-algebras and the real numbers viewed as a one-object category form an ordinary (Grothendieck) fibration over the trivial category. The fibre over each algebra is the space of states, which has a convex structure. Since real numbers have a convex structure as well, one can make sense of convexity, concavity, or affinity of the functor that computes the entropy change along a morphism of non-commutative probability spaces. The references for fibrations that we follow include~\cites{Ha16,MoVa18,LoRi19}. 

\begin{defn}[Discrete fibration]{defn:discretefibration}
A functor $\mathcal{E}\xrightarrow{\pi}\mathcal{X}$ is a \define{discrete fibration} %
 iff for each morphism $x\xrightarrow{f}y$ in $\mathcal{X}$ and for each object $v$ in $\mathcal{E}$ such that $\pi(v)=y$, there exists a unique morphism $u\xrightarrow{\beta}v$ such that $\pi(\beta)=f$. A morphism $u\xrightarrow{\beta}v$ such that $\pi(\beta)=f$ is called a \define{lift} of $f$. 
\end{defn}


\begin{exa}[The discrete fibration of non-commutative probability spaces]{ex:NCfinprobopfibration}
The functor $\pi:\NCFP\to\fdCAlg$, which sends $(\mA,\omega)$ to $\mA$ and $(\mB,\xi)\xrightarrow{f}(\mA,\omega)$ to $\mB\xrightarrow{f}\mA$, is a discrete fibration. 
Indeed, given $\w\in\mS(\mA)$ and $\mB\xrightarrow{f}\mA$, the unique lift is $f$ itself together with the state on $\mB$ given by $\xi=\w\circ f$. 
Similarly, the functor $\FinProb^{\op}\to\FinSet^{\op}$ sending a probability space $(X,p)$ to $X$ and a probability-preserving function to the underlying function between sets is a discrete fibration. 
\end{exa}

\begin{defn}[Cartesian morphisms and fibrations]{defn:cartesian}
Let $\mathcal{E}$ and $\mathcal{X}$ be two categories and let $\mathcal{E}\xrightarrow{\pi}\mathcal{X}$ be a functor. A morphism $u\xrightarrow{\beta}v$ in $\mathcal{E}$ is \define{cartesian} iff for any morphism $x\xrightarrow{f}\pi(u)$ in $\mathcal{X}$ and any morphism $w\xrightarrow{\gamma}v$ in $\mathcal{E}$ such that $\pi(\beta)\circ f=\pi(\gamma)$, there exists a unique morphism $w\xrightarrow{\alpha}u$ in $\mathcal{E}$ such that 
$
\pi(\alpha)=f
$
and 
$\beta\circ\alpha=\gamma$.
Let $\mathcal{E}_{x}$ be the subcategory of $\mathcal{E}$ consisting of the objects $u$ in $\mathcal{E}$ such that $\pi(u)=x$ and $\pi(\beta)=\id_{x}$ for all morphisms $u\xrightarrow{\beta}v$ with $\pi(u)=x=\pi(v)$. The category $\mathcal{E}_{x}$ is called the \define{fibre} of $\pi$ over $x$ and the morphisms in $\mathcal{E}_{x}$ are called \define{vertical morphisms} of $\pi$ over $x$. 
Given a morphism $x\xrightarrow{f}y$ in $\mathcal{X}$ and an object $v$ in $\mathcal{E}_{y}$, a \define{cartesian lifting of $f$ with target $v$} is a cartesian morphism $u\xrightarrow{\beta}v$ such that $\pi(\beta)=f$.
A functor $\pi:\mathcal{E}\to\mathcal{X}$ is a \define{fibration} iff for any morphism  $x\xrightarrow{f}y$ in $\mathcal{X}$ and an object $v$ in $\mathcal{E}_{y}$, a cartesian lifting exists. When $\pi$ is a fibration, 
$\mathcal{X}$ is called the \define{base}. A fibration for which a cartesian lifting has been chosen for every pair $(f,v)$, with $f$ a morphism in $\mathcal{X}$ and $v$ an object in $\mathcal{E}_{y}$, is called a \define{cloven} fibration.
\end{defn}

\begin{lem}[The reindexing functor]{lem:reindexingfunctor}
Let $\mathcal{E}\xrightarrow{\pi}\mathcal{X}$ be a cloven fibration 
and let $f^*(v)\xrightarrow{f_{v}}v$ be the choice of cartesian lifting of $x\xrightarrow{f}y$ with target $v$. These data determine a canonical functor $\mathcal{E}_{x}\xleftarrow{f^*}\mathcal{E}_{y}$ sending $v$ to $f^*(v)$. For each vertical morphism $w\xrightarrow{\kappa}v$ in $\mathcal{E}_{y}$, let $f^*(w)\xrightarrow{f^*(\kappa)}f^*(v)$ be the unique morphism in $\mathcal{E}_{x}$ obtained by the universal property of $f_{v}$ being cartesian. 
Then $f^*$ defines a functor, called the \define{reindexing functor} associated to $f$. 
\end{lem}

\bprf
This is a standard fact that follows from the uniqueness in the universal property of cartesian morphisms. The details are left as an exercise. 
\eprf

To incorporate convex structures on our main examples, we define (strict) convex categories, affine functors, and fibrewise convex structures on fibrations. 
The following definition of a convex object is an internalization of the algebraic definition of a convex space~\cites{vNMo43,St49,Ne70,Gu73,Sw74,Gu79,Fr09,Fl80}. 

\begin{defn}[Convex category]{defn:convexcategory}
Given two numbers $\l,\mu\in[0,1]$ set
\[
\l\llcorner\mu:=\l\mu 
\quad\text{ and }\quad
\l\lrcorner\mu
:=\begin{cases}
\frac{\l (1-\mu)}{1-\l\mu} & \mbox{ if } \l \mu \ne 1 \\
\text{arbitrary} &\mbox{ if } \l = \mu = 1
\end{cases}, 
\]  
where ``arbitrary'' means that one can assign any value to the
quantity.
A \define{convex category} (or more generally a \define{convex object} in some cartesian monoidal category) is a category $\mC$ (object) together with a family of functors $F_{\l}:\mC\times\mC\to\mC$ (morphisms) indexed
by $\l\in[0,1]$ such that 
\[
\xy0;/r.20pc/:
(-12.5,0)*+{\mC\times\mC}="CC";
(12.5,0)*+{\mC}="C";
{\ar@/^1.25pc/"CC";"C"^{F_{0}}};
{\ar@/_1.25pc/"CC";"C"_{\pi_{2}}};
\endxy
\quad,\quad
\xy0;/r.20pc/:
(-10,-7.5)*+{\mC}="C1";
(0,7.5)*+{\mC\times \mC}="CC";
(10,-7.5)*+{\mC}="C2";
{\ar"C1";"CC"^{\D}};
{\ar"CC";"C2"^{F_{\l}}};
{\ar"C1";"C2"_{\id_{\mC}}};
\endxy
\quad,\quad
\xy0;/r.20pc/:
(-10,7.5)*+{\mC\times\mC}="CC1";
(10,7.5)*+{\mC\times\mC}="CC2";
(0,-7.5)*+{\mC}="C";
{\ar"CC1";"CC2"^{\g}};
{\ar"CC1";"C"_(0.4){F_{\l}}};
{\ar"CC2";"C"^(0.4){F_{1-\l}}};
\endxy
\;,\;\text{ and }\;
\xy0;/r.20pc/:
(-17.5,7.5)*+{\mC\times\mC\times\mC}="1";
(17.5,7.5)*+{\mC\times\mC}="2";
(-17.5,-7.5)*+{\mC\times\mC}="3";
(17.5,-7.5)*+{\mC}="4";
{\ar "1";"2"^(0.55){F_{\mu} \times \id_{\mC}}};
{\ar "1";"3"_{\id_{\mC} \times F_{\l\lrcorner\mu}}};
{\ar "2";"4"^{F_{\l}}};
{\ar "3";"4"_{F_{\l\llcorner\mu}}};
\endxy
\]
commute for all $\l,\mu\in[0,1]$ (see Definition~\ref{defn:categorynotation} for notation). 
The notation $\l x + (1-\l) y := F_{\l} (x,y)$ will be implemented on occasion. 
\end{defn}

\begin{exa}[Examples of convex categories]{ex:convexobjects}
{\color{white}{hello!}}
\begin{enumerate}[(a)]
\itemsep0pt
\item
\label{item:statesconvexcategory}
Every convex set is a convex category when viewed as a discrete category. 
In particular, $\mS(\mA)$, the set of states on a $C^*$-algebra $\mA$, is a convex category. 
\item
The convex combination of real numbers turns $\mathbb{B}\R$ into a convex category. 
If $\R_{\ge0}:=\{r\in\R\,:\,r\ge0\}$, then $\B\R_{\ge0}$ is also a convex category. 
\end{enumerate}
Note, however, that the convex categories of BFL~\cite{BFL} are \emph{not} examples of Definition~\ref{defn:convexcategory} (cf.\ Remark~\ref{rmk:finprobnotaconvexcat}). 
\end{exa}

\begin{defn}[Affine functors]{defn:affinemap}
An \define{affine functor}
from one convex category $(\mC,\{F_{\l}\})$ to another one $(\mD,\{G_{\l}\})$ is
a functor $S:\mC\to\mD$ such that 
\[
\xy0;/r.20pc/:
(-12.5,7.5)*+{\mC\times \mC}="CC";
(12.5,7.5)*+{\mD\times\mD}="DD";
(-12.5,-7.5)*+{\mC}="C";
(12.5,-7.5)*+{\mD}="D";
{\ar"CC";"DD"^{S\times S}};
{\ar"CC";"C"_{F_{\l}}};
{\ar"C";"D"_{S}};
{\ar"DD";"D"^{G_{\l}}};
\endxy
\]
commutes for all $\l \in [0,1]$.
\end{defn}

\begin{exa}[The pullback of states is an affine functor]{ex:pullbackstates}
Let $\mB\xrightarrow{f}\mA$ be a $*$-homomorphism between $C^*$-algebras. Then the pullback $\mS(\mA)\xrightarrow{\mS(f)}\mS(\mB)$, sending $\w$ to $\w\circ f$, is an affine functor (cf.\ Example~\ref{ex:convexobjects}~(\ref{item:statesconvexcategory})) since
\[
\big(\l\w+(1-\l)\xi\big)\circ f=\l(\w\circ f)+(1-\l)(\xi\circ f)
\qquad\forall\;\l\in[0,1],\;\w,\xi\in\mS(\mA).
\] 
\end{exa}

\begin{exa}[Entropy is almost affine]{ex:entropyalmostaffine}
Given $\mB\xrightarrow{f}\mA$, the assignment $\mS(\mA)\xrightarrow{S_{f}}\B\R$ sending $\w$ to $S_{f}(\w)$ from Definition~\ref{defn:entropychange} is \emph{not} affine. However, the inequality
\[
S_{f}\big(\l\w+(1-\l)\xi\big)\ge\l S_{f}(\w)+(1-\l)S_{f}(\xi)
\]
holds as a corollary of the work of Lieb and Ruskai~\cite[Theorem~1]{LiRu73} and Lindblad~\cite[Lemma~3]{Li73}. Nevertheless, and more importantly for our characterization theorem, equality \emph{does} hold when $\w\perp\xi$ and $(\w\circ f)\perp(\xi\circ f)$. The proof of this will be given in Proposition~\ref{prop:entropydifferenceisnice}. 
\end{exa}

\begin{defn}[Fibrewise convex structures]{defn:fibredconvexstructures}
A \define{fibrewise convex structure} on a fibration $\mathcal{E}\xrightarrow{\pi}\mathcal{X}$ is a cloven fibration where each fibre is a convex category and each reindexing functor $\mathcal{E}_{x}\xleftarrow{f^*}\mathcal{E}_{y}$ (as described in Lemma~\ref{lem:reindexingfunctor}) is an affine functor. A cloven fibration equipped with a fibrewise convex structure is called a \define{fibrewise convex fibration}. 
\end{defn}

\begin{exa}[Examples of fibrewise convex structures]{exa:fibrewiseconvex}
{\color{white}{hello!}}
\begin{enumerate}[(a)]
\itemsep0pt
\item
The discrete fibration $\NCFP\to\fdCAlg$ has $\mS(\mA)$ as the fibre over each $C^*$-algebra $\mA$. The set of states $\mS(\mA)$ on a $C^*$-algebra $\mA$ has a natural convex structure. Furthermore, each $*$-homomorphism $\mB\xrightarrow{f}\mA$ has the pullback $\mS(\mB)\xleftarrow{\mS(f)}\mS(\mA)$ as its reindexing functor. This functor is affine, as discussed in Example~\ref{ex:pullbackstates}. 
\item
By a similar argument, $\FinProb^\op\to\FinSet^{\op}$ has a natural fibrewise convex structure coming from the convex combination of probability measures and the fact that the pushforward of measures is linear. The fibre over a finite set $X$ is isomorphic to the standard simplex $\Delta^{|X|-1}:=\big\{(p_{1},\dots,p_{|X|})\in\R^{|X|}_{\ge0}\;:\;\sum_{i=1}^{|X|}p_{i}=1\big\}$. 
\item
The fibration $\B\R\to\uline{\mathbf{1}}$ has a convex structure on the only fiber $\B\R$ over the single object in the base, as described in Example~\ref{ex:convexobjects}.
\end{enumerate}
\end{exa}

\begin{defn}[Morphisms of fibrations]{defn:fibredfunctor}
Let $\mathcal{E}\xrightarrow{\pi}\mathcal{X}$ and $\mathcal{F}\xrightarrow{\rho}\mathcal{Y}$ be fibrations. A \define{fibred functor}%
\footnote{Our terminology differs from that of~\cite{MoVa18}, who use `functor' when the base category is fixed ($\phi=\id$) and `1-cell' for when the base category changes.}
 from $\pi$ to $\rho$ is a pair of functors $\mathcal{E}\xrightarrow{\Phi}\mathcal{F}$ and $\mathcal{X}\xrightarrow{\phi}\mathcal{Y}$ such that 
\[
\xy0;/r.20pc/:
(-10,7.5)*+{\mathcal{E}}="E";
(10,7.5)*+{\mathcal{F}}="F";
(-10,-7.5)*+{\mathcal{X}}="X";
(10,-7.5)*+{\mathcal{Y}}="Y";
{\ar"E";"F"^{\Phi}};
{\ar"E";"X"_{\pi}};
{\ar"X";"Y"_{\phi}};
{\ar"F";"Y"^{\rho}};
\endxy
\]
commutes and such that $\Phi(\beta)$ is cartesian for every cartesian $\beta$. 
\end{defn}

\begin{rmk}[Fibrewise convex structures as internal convex objects]{rmk:convexcombosofmorphisms}
One can equivalently define a fibrewise convex structure as an internal convex object in the category of fibrations over a fixed based, analogous to the fibrewise monoidal structure in~\cite[Section~3.1]{MoVa18}. 

Briefly, a convex object $\mathcal{E}\xrightarrow{\pi}\mathcal{X}$ in the category of fibrations over a fixed based $\mathcal{X}$ provides the data of a family of fibred functors $F_{\l}:\mathcal{E}\times_{\pi}\mathcal{E}\to\mathcal{E}$ with a fixed based, 
where
$\mathcal{E}\times_{\pi}\mathcal{E}$ is the (strict) pullback. 
The functors $F_{\l}$ define a convex category structure for every fibre $\mathcal{E}_{x}$. In addition, they also provide an assignment on morphisms since a pair $(t\xrightarrow{\alpha}u,v\xrightarrow{\beta}w)$ over $x\xrightarrow{f}y$ gets sent to 
\[
\l t+(1-\l)v\xrightarrow{\l\alpha+(1-\l)\beta\equiv F_{\l}(\a,\b)}\l u+(1-\l)w
\] 
over $x\xrightarrow{f}y$. This assignment guarantees that the associated reindexing functor $\mathcal{E}_{x}\xleftarrow{f^*}\mathcal{E}_{y}$ from Lemma~\ref{lem:reindexingfunctor} can be chosen to be affine as in Definition~\ref{defn:fibredconvexstructures}. Indeed, if one chooses cartesian liftings $f^{*}(u)\xrightarrow{f_{u}}u$ and $f^{*}(v)\xrightarrow{f_{v}}v$ of $u$ and $v$ over $x\xrightarrow{f}y$, respectively, then  
\[
\l f^{*}(u)+(1-\l)f^{*}(v)\xrightarrow{\l f_{u}+(1-\l)f_{v}}\l u+(1-\l)v
\]
can be taken as the lift of $\l u+(1-\l)v$ over $f$. 

For example, in the fibrewise convex fibration $\NCFP\to\fdCAlg$, if $(\mB,\eta)\xrightarrow{g}(\mA,\w)$ and $(\mB,\z)\xrightarrow{h}(\mA,\xi)$ are two morphisms over $\mB\xrightarrow{f}\mA$, then $g=h=f$ and their convex combination, $\l g+(1-\l)h$, is just $f$. In the fibrewise convex fibration $\B\R\to\uline{\mathbf{1}}$, the convex combination of objects in the fibre is trivial, while the convex combination of morphisms (elements in $\R$) is the usual convex combination of real numbers. 
\end{rmk}

\begin{defn}[Convergence in $\NCFP$]{defn:NCFP}
A sequence $\N\ni n\mapsto\big((\mB_{n},\xi_{n})\xrightarrow{f_{n}}(\mA_{n},\w_{n})\big)$ \define{converges to} $(\mA,\xi)\xrightarrow{f}(\mB,\w)$ in the category $\NCFP$ iff there exists an $N\in\N$ such that 
$\mA_{n}=\mA$, $\mB_{n}=\mB$, $f_{n}=f$ for all $n\in\N$, 
$\lim_{n\to\infty}\w_{n}=\w$, and $\lim_{n\to\infty}\xi_{n}=\xi$, 
where the last two limits are with respect to the standard topologies on the state spaces $\mS(\mA)$ and $\mS(\mB)$, respectively.%
\end{defn}

\begin{rmk}[Justifying the definition of convergence of sequences in $\NCFP$]{rmk:justifyingcontinuity}
The definition of convergence of a sequence of morphisms in $\NCFP$ is motivated by the one in $\FinProb$ from~\cite[page 4]{BFL}. However, some justification regarding why the morphisms are assumed to stabilize, i.e.\ are equal after some $N\in\N$, is needed.
 
In the case of $\FinProb$, a sequence $(X_{n},p_{n})\xrightarrow{f_{n}}(Y_{n},q_{n})$ \define{converges to} $(X,p)\xrightarrow{f}(Y,q)$ iff the sets $X_{n}, Y_{n}$ and the underlying set functions $f_{n}$ stabilize after a finite natural number in the sequence and $\lim_{n\to\infty}p_{n}=p$ and $\lim_{n\to\infty}q_{n}=q$. The sets must stabilize because their associated simplices of probability distributions are distinct and the cardinality of the set dictates which simplex one is using for the space of probability distributions. The functions must stabilize because the set of functions between two finite sets is also a finite set, which has the discrete topology. However, the probability distributions $p_{n}$ on $X$ and $q_{n}$ on $Y$ may continue to vary as long as they converge to $p$ and $q$ in the topology associated with the simplices $\Delta^{|X|-1}$ and $\Delta^{|Y|-1}$. 

In the case of $C^*$-algebras, the collection $\mathrm{hom}(\mB,\mA)$ of (unital) $*$-homomorphisms from $\mB$ to $\mA$ is \emph{not} just a discrete set since the collection of unitary matrices has a non-trivial topology. 
Nevertheless, one can assume the $f_{n}$ eventually stabilize. To see this, it suffices to assume $\mA=\bigoplus_{x\in X}\mM_{m_{x}}$ and $\mB=\bigoplus_{y\in Y}\mM_{n_{y}}$ for some finite sets $X$ and $Y$ and $m_{x},n_{y}\in\N$. In this case, a $*$-homomorphism $\mB\xrightarrow{f}\mA$ is described by its multiplicities and by a unitary as in Lemma~\ref{lem:ptracedirectsums}. 
The multiplicities entail the constraint $m_{x}=\sum_{y\in Y}c_{xy}n_{y}$, but there could be several such multiplicities satisfying these constraints. Indeed, if%
\[
s_{x}:=\Bigg|\Bigg\{Y\ni y\mapsto c_{xy}\in\Z_{\ge0}\;:\;m_{x}=\sum_{y\in Y}c_{xy}n_{y}\Bigg\}\Bigg|
\]
denotes the number of such solutions, then the number of connected components in $\mathrm{hom}(\mB,\mA)$ is $s:=\prod_{x\in X}s_{x}$ (for example, if $\mB=\mM_{n}$ is a matrix algebra, there is only one such component). 
Hence, a sequence of $*$-homomorphisms converging to another one must necessarily have multiplicities that stabilize. Within such a component, since $\w\circ f=(\w\circ\mathrm{Ad}_{U})\circ(\mathrm{Ad}_{U^{\dag}}\circ f)$ for every unitary $U$, one can always choose $f$ to be of the form 
\[
\bigoplus_{y\in Y}\mM_{n_{y}}\ni \bigoplus_{y\in Y}B_{y}\mapsto\bigoplus_{x\in X}\Bigg(\bigboxplus_{y\in Y}
\mathds{1}_{c_{yx}}\otimes B_{y}
\Bigg)
\]
by conjugating with some appropriate unitary $U$ (cf.\ Lemma~\ref{lem:ptracedirectsums}). This unitary can then be transferred to the state. 

Therefore, it suffices to assume the algebras and $*$-homomorphisms stabilize in a convergent sequence, but not necessarily the states. 
\end{rmk}

\begin{defn}[Continuous fibred functors]{defn:contfibredfunctor}
A \define{continuous fibred functor} from $\NCFP\to\fdCAlg$ to $\B\R\to\uline{\mathbf{1}}$ is a fibred functor $H$  
such that to every sequence $\N\ni n\mapsto\big((\mB_{n},\xi_{n})\xrightarrow{f_{n}}(\mA_{n},\w_{n})\big)$ converging to $(\mA,\xi)\xrightarrow{f}(\mB,\w)$ in the category $\NCFP$, 
\[
\lim_{n\to\infty}H\left((\mB_{n},\xi_{n})\xrightarrow{f_{n}}(\mA_{n},\w_{n})\right)=H\left((\mB,\xi)\xrightarrow{f}(\mA,\w)\right),
\]
where the convergence is for a sequence of real numbers. 
\end{defn}

\begin{notat}[The function $H_{f}:\mS(\mA)\to\R$]{not:Hf}
For a fibred functor $H:\NCFP\to\B\R$, set
\[
H_{f}(\w):=H\left((\mB,\xi)\xrightarrow{f}(\mA,\w)\right)
\]
for the image of $H$ along a morphism $f$ in $\NCFP$. 
For a fixed $*$-homomorphism $\mB\xrightarrow{f}\mA$, this defines a function $H_{f}:\mS(\mA)\to\R$. 
\end{notat}

The next definition is the appropriate quantum generalization of the affinity condition used by BFL in their characterization of Shannon entropy~\cite{BFL}. Why this is so will be explained towards the end of this section as well as Proposition~\ref{lem:externalaffineorthogonalaffine} and Remark~\ref{rmk:externalignoresinternal}. 

\begin{defn}[Orthogonally affine fibred functor]{defn:orthogonallyaffinefibredfunctor}
A fibred functor $H$ from $\NCFP\to\fdCAlg$ to $\B\R\to\uline{\mathbf{1}}$ is \define{orthogonally affine} iff to each pair of $C^*$-algebras $\mB$ and $\mA$, each pair of mutually orthogonal states $\w,\xi\in\mS(\mA)$, and each $*$-homomorphism $\mB\xrightarrow{f}\mA$ such that $(\w\circ f)\perp(\xi\circ f)$,  
\[
H_{f}\big(\l\w+(1-\l)\xi\big)=\l H_{f}(\w)+(1-\l)H_{f}(\xi)\qquad\forall\;\l\in[0,1].
\]
\end{defn}

\begin{prop}[Entropy difference is continuous and orthogonally affine]{prop:entropydifferenceisnice}
The entropy change functor 
from Definition~\ref{defn:entropychange} is a continuous and orthogonally affine fibred functor. In fact, if for any $C^*$-algebra $\mA$ and any pair $\w,\xi$ of mutually orthogonal states on $\mA$, a $*$-homomorphism $\mB\xrightarrow{f}\mA$ preserves the orthogonality $\omega\perp\xi$ if and only if
\[
S_{f}\big(\l\w+(1-\l)\xi\big)=\l S_{f}(\w)+(1-\l)S_{f}(\xi)\qquad\forall\;\l\in[0,1].
\]
\end{prop}

Before proving this, we introduce a shorthand for the deviation from $S_{f}$ being affine on the states $\omega$ and $\xi$. The name for this deviation is motivated by~\cite[Section~12.1.1]{NiCh11}. 

\begin{defn}[The Holevo information change along a morphism]{defn:entropydeviation}
The \define{Holevo information change along} a $*$-homomorphism $\mB\xrightarrow{f}\mA$ associated to $\omega,\xi\in\mS(\mA)$ and $\l\in[0,1]$ is the number
\[
\chi_{f}(\l;\omega,\xi):=S_{f}\big(\l\omega+(1-\l)\xi\big)-\l S_{f}(\w)-(1-\l)S_{f}(\xi). 
\]
\end{defn}

Proposition~\ref{prop:entropydifferenceisnice} says, in particular, that this deviation vanishes when $\w\perp\xi$ and $(\w\circ f)\perp(\xi\circ f)$. 


\bprf[Proof of Proposition~\ref{prop:entropydifferenceisnice}]
Continuity of the entropy change follows from continuity of the von~Neumann entropy~\cite[Section~11.3]{NiCh11},~\cite{Fa73}. To prove the statement regarding orthogonal affinity, suppose $\w\perp\xi$. 
Let $\w':=\w\circ f$ and $\xi':=\xi\circ f$. 
If $f$ preserves the mutual orthogonality, then $\w'\perp\xi'$ and 
\be
\begin{split}
\chi_{f}(\l;\w,\xi)&=S\big(\l\w+(1-\l)\xi\big)-S\big(\l\w'+(1-\l)\xi'\big)-\l S_{f}(\w)-(1-\l)S_{f}(\xi)\\
&\overset{\text{Lem~\ref{lem:derivationpropertyonorthogonal}}}{=\joinrel=\joinrel=\joinrel=\joinrel=}
S(\l,1-\l)+\l S(\w)+(1-\l)S(\xi)\\
&\qquad\;\;-S(\l,1-\l)-\l S(\w')-(1-\l)S(\xi')\\
&\qquad\qquad\qquad\qquad\;-\l S_{f}(\w)-(1-\l)S_{f}(\xi)\\
&=0,
\end{split}
\ee
where $S(\l,1-\l)$ is the Shannon entropy of the probability $(\l,1-\l)$ on a two element set. 
Conversely, suppose $\chi_{f}(\l;\w,\xi)=0$. 
Since $\w\perp\xi$, a similar calculation gives
\be
0=\chi_{f}(\l;\w,\xi)\\
\overset{\text{Lem~\ref{lem:derivationpropertyonorthogonal}}}{=\joinrel=\joinrel=\joinrel=\joinrel=}S(\l,1\!-\!\l)+\l S(\w')+(1\!-\!\l)S(\xi')-S\big(\l\w'\!+\!(1\!-\!\l)\xi'\big),
\ee
which gives 
$\w'\perp\xi'$ 
by the `only if' part of Lemma~\ref{lem:derivationpropertyonorthogonal}.
\eprf


In the last part of this section, we recall the convex combinations and affine functors introduced by BFL~\cite{BFL}. 
By the next section, we will have enough facts to relate BFL's definition to ours. 

\begin{defn}[An external convex structure on $\FinProb$]{defn:externalconvexfinprob}
For every $\l\in[0,1],$ define the convex sum $F_{\l}$ on objects of $\FinProb$ by
\[
\l(X,p)\oplus(1-\l)(Y,q):=\big(X\amalg Y,\l p\oplus(1-\l)q\big),
\]
where $\l p\oplus(1-\l)q$ is defined in Example~\ref{ex:BFLexternal}. 
The convex sum of morphisms $(X,p)\xrightarrow{\phi}(X',p')$
and $(Y,q)\xrightarrow{\psi}(Y',q')$ is defined to be the disjoint union $\phi\amalg\psi$ as in Example~\ref{ex:BFLexternal}. 
The collection of functors $\{F_{\l}\}_{\l\in[0,1]}$ is called the \define{external convex structure} on $\FinProb$. 
\end{defn}

The motivation for calling this an \emph{external} convex structure comes from the distinction between internal and external monoidal fibrations~\cite[Section~3.1]{MoVa18}, as will be explained shortly. 


\begin{rmk}[The external convex structure on $\FinProb$ does not give a convex category]{rmk:finprobnotaconvexcat}
$\FinProb$ with this family of functors is \emph{not} a convex category in the sense of Definition~\ref{defn:convexcategory}. It is, however, a \emph{weak} convex category (called a convex category in~\cite[Chapter~4]{Pa16}). 
\end{rmk}

A completely analogous definition can be made for the fibration $\NCFP\to\fdCAlg$ using the (external) direct sum of $C^*$-algebras. 

\begin{defn}[An external convex structure on $\NCFP$]{defn:externalconvexNCFP}
For every $\l\in[0,1],$ define the convex sum $F_{\l}$ on objects of $\NCFP$ by $\l(\mA,\omega)\oplus(1-\l)(\mB,\xi):=\big(\mA\oplus\mB,\l\omega\oplus(1-\l)\xi\big)$, where $\big(\l\omega\oplus(1-\l)\xi\big)(A\oplus B):=\l\omega(A)+(1-\l)\xi(B)$ for all $A\in\mA$, $B\in\mB$. The convex sum of morphisms is the direct sum. 
\end{defn}

This convex structure on $\NCFP$ restricts to the one on $\FinProb$ on the subcategory of commutative $C^*$-algebras since $\C^{X\amalg Y}\cong\C^{X}\oplus\C^{Y}$. 

\begin{defn}[Externally affine functor]{defn:externallyaffinefunctor}
A functor $H:\NCFP\to\B\R$ is \define{externally affine} iff
\[
H\big(\l f\oplus(1-\l)g\big)=\l H(f)+(1-\l)H(g)
\]
for all morphisms $f,g$ in $\NCFP$ and all $\l\in[0,1]$. 
\end{defn}

\begin{exa}[Examples of externally affine functors]{ex:externallyaffine}
{\color{white}{you found me!}}
\begin{enumerate}[(a)]
\itemsep0pt
\item
The difference of Shannon entropies studied by BFL~\cite{BFL} is a continuous externally affine functor $\FinProb\to\B\R$. In fact, it is characterized as the unique one whose image always lands in $\B\R_{\ge0}$ (cf.\ Theorem~\ref{thm:BFL}). 
\item
An example of a continuous externally affine functor $S:\NCFP\to\B\R$ is the difference of Segal entropies from Definition~\ref{defn:entropychange}. 
\item
\label{item:quantumShannondifference}
If $f:(\mB,\xi)\xrightarrow{f}(\mA,\omega)$ is as in Lemma~\ref{lem:ptracedirectsums}, then  
$K_{f}(\w):=S(p)-S(q),$
the difference of the Shannon entropies associated to the probability distributions, %
defines a continuous externally affine functor $K:\NCFP\to\B\R$.
\end{enumerate}
Notice that both $K$ and $S$ agree with the Shannon entropy difference on the subcategory of commutative algebras, yet they are \emph{not} proportional.%
\footnote{The existence of these two distinct continuous (externally) affine functors illustrates that continuous affine functors $\NCFP\to\B\R$ are not characterized by their values on $\FinProb^{\op}$ (when viewed as a subcategory of $\NCFP$). In particular, this condition does not characterize the von~Neumann entropy difference. This answers a question of John Baez in the negative~\cite{BaezIII} (see specifically the original post as well as the post on June 7, 2011 at 8:12 AM).}
\end{exa}

For reference, we recall BFL's characterization theorem~\cite{BFL}. 

\begin{theo}[BFL's functorial characterization of the Shannon entropy]{thm:BFL}
If $H:\FinProb\to\B\R_{\ge0}$ is a continuous externally affine functor, then there exists a constant $c\ge0$ such that 
$H_{\phi}(p)=c\big(S(p)-S(q)\big)$ 
for every probability-preserving function $(X,p)\xrightarrow{\phi}(Y,q)$. 
\end{theo}

Without reference to the entropy formulas from Definition~\ref{defn:entropychange}, we will relate internal and external affinity in Proposition~\ref{lem:externalaffineorthogonalaffine} after developing some general results. 

\section[Characterizing entropy]{Characterizing entropy}
\label{sec:classifyentropy}

This section contains our main result, Theorem~\ref{thm:abstractentropy}, which is a functorial characterization of the entropy difference in the non-commutative setting. 
Continuity and orthogonal affinity alone are not quite enough to characterize the von~Neumann entropy difference, though they come quite close. 
By Lemma~\ref{lem:entropychangecommutative}, we cannot assume that $S_{f}(\w)\ge0$ for all $*$-homomorphisms $f$ and states $\w$ on the codomain of $f$, since this inequality fails for non-commutative $C^*$-algebras. 

We propose a close replacement, namely $S_{\mA}(\w)\ge0$ for all states $\w\in\mS(\mA)$, with equality on pure states, for all $C^*$-algebras $\mA$. While this may sound quite different, this assumption is a consequence of BFL's assumption $S_{f}(\w)\ge0$ on \emph{commutative} $C^*$-algebras. 
Furthermore, in Proposition~\ref{prop:positivitycommutative}, we prove that the non-negativity of entropy difference for \emph{commutative} $C^*$-algebras is a \emph{consequence} of the fact that state-preserving $*$-homomorphisms between commutative $C^*$-algebras always have \emph{disintegrations}. 
More generally, we show that the existence of disintegrations (with non-commutative probability spaces included) implies the non-negativity of entropy difference.  

\begin{notat}[$!_{\mA}$ and $H_{\mA}$]{not:uniqueHA}
If $\mA$ is a $C^*$-algebra, then $\C\xrightarrow{!_{\mA}}\mA$ will always refer to the unique (unital) $*$-homomorphism. 
If $H:\NCFP\to\B\R$ is a functor, set $H_{\mA}:=H_{!_{\mA}}$. Also, $\FinProb^{\op}$ will be viewed as the full subcategory of $\NCFP$ consisting of commutative probability spaces.
\end{notat}

\begin{lem}[$H$ is a coboundary]{lem:Hcoboundary}
Given any $*$-homomorphism $\mB\xrightarrow{f}\mA$ and a state $\mA\xstoch{\omega}\C$, any functor $H:\NCFP\to\B\R$ satisfies
\[
H_{f}(\w)=H_{\mA}(\w)-H_{\mB}(\w\circ f).
\]
\end{lem}

\bprf
This follows from 
$\C$ being an initial object in $\fdCAlg$.
\eprf

\begin{lem}[Non-negativity of $H_{f}$ implies vanishing of $H_{\mA}$ on pure states]{lem:BFLimplieszeroonpure}
Let $H:\FinProb^{\op}\to\B\R$ be a functor satisfying $H_{f}(\w)\ge0$ for all $\w\in\mS(\mA)$ and $*$-homomorphisms $\mB\xrightarrow{f}\mA$ between commutative $C^*$-algebras. %
\begin{enumerate}
\itemsep0pt
\item
If $f$ has a left or right inverse, then $H_{f}(\omega)=0$ for all $\omega\in\mS(\mA)$. 
\item
$H_{\mA}(\w)\ge0$ for all states $\w\in\mS(\mA)$, with equality on all pure states.
\end{enumerate}
\end{lem}

\bprf
{\color{white}{you found me!}}
\begin{enumerate}
\itemsep0pt
\item
Suppose $f$ has a right inverse $\mA\xrightarrow{g}\mB$. Then functoriality of $H$ implies $0=H_{\id_{\mA}}(\w)=H_{g}(\w\circ f)+H_{f}(\w)$ by Lemma~\ref{lem:Hcoboundary}. Since each term is non-negative by assumption, $H_{f}(\w)\ge0$. A similar calculation proves the same inequality if $f$ has a left inverse. 
\item
First, $H_{\mA}(\w)=H_{!_{\mA}}(\w)\ge0$ by assumption.  
By invariance of $H$ under $*$-isomorphisms, it suffices to take $\mA=\C^{X}$, with $X$ a finite set. Any pure state $\xi$ on $\C^{X}$ is necessarily supported on some $x\in X$. Let $\C^{X}\xrightarrow{\pi_{x}}\C$ be the projection onto that component. Then $\pi_{x}$ pulls the unique state $1$ on $\C$ back to $\xi$ on $\C^{X}$ and the composite $\C\xrightarrow{!_{\C^{X}}}\C^{X}\xrightarrow{\pi_{x}}\C$ equals $\id_{\C}$. Thus, $H_{\C^{X}}(\xi)=0$ by the first item. \qedhere
\end{enumerate}
\eprf

A partial converse to Lemma~\ref{lem:BFLimplieszeroonpure} will illustrate that our axioms for entropy change imply those of BFL. We first prove a lemma about invariance under $*$-isomorphisms given \emph{our} axioms. The proof is quite different from the one in Lemma~\ref{lem:BFLimplieszeroonpure}, and it uses the  convex structure in a crucial way. 

\begin{lem}[$H$ is invariant under $*$-isomorphisms]{lem:Hinvariantunderiso}
Suppose $H:\NCFP\to\B\R$ is an orthogonally affine fibred functor for which $H_{\mA}(\xi)=0$ for all pure states $\xi$ on $\mA$ and all $C^*$-algebras $\mA$. If $\mB\xrightarrow{f}\mA$ is a $*$-isomorphism, then 
$H_{f}(\w)=0$ for all $\w\in\mathcal{S}(\mA)$. 
\end{lem}

\bprf
Let $\w$ be a state on $\mA$. Then there exists a convex decomposition $\w=\sum_{x\in X}p_{x}\w_{x}$ of $\w$ in terms of mutually orthogonal pure states $\w_{x}$ and a nowhere-vanishing probability measure $p$ on some finite set $X$. 
Thus, 
\be
\begin{split}
H_{f}(\w)&\underset{\text{Defn~\ref{defn:orthogonallyaffinefibredfunctor}}}{\overset{\text{Lem~\ref{lem:isopreserveperp}}}{=\joinrel=\joinrel=\joinrel=\joinrel=\joinrel=}}\sum_{x\in X}p_{x}H_{f}(\w_{x})\\
&\overset{\text{Lem~\ref{lem:Hcoboundary}}}{=\joinrel=\joinrel=\joinrel=\joinrel=}\sum_{x\in X}p_{x}\Big(H_{\mA}(\w_{x})-H_{\mB}(\w_{x}\circ f)\Big)
=0
\end{split}
\ee
since $\w_{x}\circ f$ is pure by Lemma~\ref{lem:isopreserveperp}. 
\eprf

\begin{defn}[Disintegrations on finite probability spaces]{defn:disintegration}
{\color{white}{you found me!}}

\noindent
\begin{minipage}{0.63\textwidth}
Let $(X,p)$ and $(Y,q)$ be probability spaces and let
$\phi:X\to Y$ be a probability-preserving function, i.e.\ $q=\phi\circ p$.
A \define{disintegration} of 
$(\phi,p,q)$ (or simply of $\phi$ if $p$ and $q$ are clear from context)
is a stochastic map
$Y\xstoch{\psi}X$ such that 
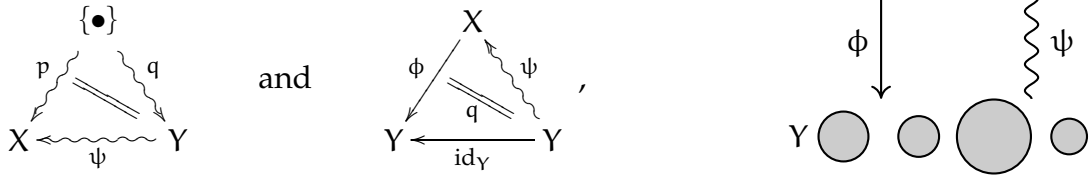
\[
\xy0;/r.25pc/:
(0,7.5)*+{\{\bullet\}}="o";
(-10,-7.5)*+{X}="X";
(10,-7.5)*+{Y}="Y";
{\ar@{~>}"o";"X"_{p}};
{\ar@{~>}"o";"Y"^{q}};
{\ar@{~>}"Y";"X"^{\psi}};
{\ar@{=}(-3,0);(5,-4.5)};
\endxy
\qquad\text{and}\qquad
\xy0;/r.25pc/:
(0,7.5)*+{X}="X";
(10,-7.5)*+{Y}="Y1";
(-10,-7.5)*+{Y}="Y2";
{\ar@{~>}"Y1";"X"_{\psi}};
{\ar"X";"Y2"_{\phi}};
{\ar"Y1";"Y2"^{\mathrm{id}_{Y}}};
{\ar@{=}(-3,0);(5,-4.5)_{q}};
\endxy
\;,
\]
\end{minipage}
\;
\begin{minipage}{0.30\textwidth}
\centering
    \begin{tikzpicture}[decoration=snake]
\node at (-1.60,3.25) {$X$};
\node at (-1.60,0) {$Y$};
\draw[thick,fill=black,fill opacity=0.2] (-1,3.75) circle (1.0ex);
\draw[thick,fill=black,fill opacity=0.2] (-1,3.25) circle (0.5ex);
\draw[thick,fill=black,fill opacity=0.2] (-1,2.75) circle (1.25ex);
\draw[thick,fill=black,fill opacity=0.2] (0,4.5) circle (0.25ex);
\draw[thick,fill=black,fill opacity=0.2] (0,4.0) circle (0.75ex);
\draw[thick,fill=black,fill opacity=0.2] (0,3.5) circle (1.0ex);
\draw[thick,fill=black,fill opacity=0.2] (0,3) circle (0.5ex);
\draw[thick,fill=black,fill opacity=0.2] (1,4) circle (1.75ex); 
\draw[thick,fill=black,fill opacity=0.2] (1,3.45) circle (0.75ex); 
\draw[thick,fill=black,fill opacity=0.2] (1,3) circle (1.0ex); 
\draw[thick,fill=black,fill opacity=0.2] (1,2.5) circle (1.25ex); 
\draw[thick,fill=black,fill opacity=0.2] (2,3.75) circle (0.75ex); 
\draw[thick,fill=black,fill opacity=0.2] (2,3.25) circle (0.25ex); 
\draw[thick,fill=black,fill opacity=0.2] (2,2.75) circle (0.5ex); 
\draw[thick,fill=black,fill opacity=0.2] (2,2.25) circle (0.75ex); 
\draw[-{>[scale=2.5,
          length=2,
          width=3]},thick] (-0.5,2.0) -- node[left]{$\phi$} (-0.5,0.5);
\draw[-{>[scale=2.5,
          length=2,
          width=3]},decorate,thick] (1.5,0.5) -- node[right,xshift=0.1cm]{$\psi$} (1.5,2.0);
\draw[thick,fill=black,fill opacity=0.2] (-1,0) circle (1.677ex);
\draw[thick,fill=black,fill opacity=0.2] (0,0) circle (1.3693ex);
\draw[thick,fill=black,fill opacity=0.2] (1,0) circle (2.48746ex);
\draw[thick,fill=black,fill opacity=0.2] (2,0) circle (1.19895ex);\end{tikzpicture}
\end{minipage}

\noindent
the latter diagram signifying commutativity $q$-a.e.%
\footnote{The cartoon depicts probability measures as collections of water droplets with total volume 1. The map $\phi$ combines water droplets and preserves the volume~\cite{Gr14}, while the disintegration $\psi$ splits the water droplets back into their original sizes. 
} 
Here, a \define{stochastic map} $Y\xstoch{\psi}X$ associates to each $y\in Y$ a probability measure $\psi_{y}$ on $X$. Composition of stochastic maps is defined via the Chapman--Kolmogorov equation~\cite[Section~2]{Pa17}.
\end{defn}

The main fact we will use about disintegrations on finite probability spaces is that they always exist~\cite[Theorem~5.1]{PaRu19}. 

\begin{prop}[Positivity of entropy difference on commutative $C^*$-algebras]{prop:positivitycommutative}
Suppose $H:\NCFP\to\B\R$ is an orthogonally affine fibred functor for which $H_{\mA}(\w)\ge0$ for all states $\w\in\mS(\mA)$, with equality on all pure states,  for all $C^*$-algebras $\mA$. %
Then for \emph{commutative} $C^*$-algebras $\mA$ and $\mB$, 
$H_{f}(\omega)\ge0$
for all states $\w\in\mA$ and all $*$-homomorphisms $\mB\xrightarrow{f}\mA$.  
\end{prop}

\bprf
By invariance of $H$ for $*$-isomorphisms (Lemma~\ref{lem:Hinvariantunderiso}), it suffices to assume $\mB=\C^{Y}$ and $\mA=\C^{X}$ for finite sets $X$ and $Y$. In this case, let $\w$ be represented by a probability measure $p$ on $X$,  
let $X\xrightarrow{\phi}Y$ be the function associated to $\mB\xrightarrow{f}\mA$, and let $q:=\phi\circ p$ be the pushforward measure corresponding to $\w\circ f=:\xi$ (cf.\ Example~\ref{ex:BFLexternal}). Every such probability measure is decomposed as 
$
q=\sum_{y\in Y}q_{y}\de_{y},
$
where $\de_{y}$ is the Dirac delta measure at $y$ defined by $\de_{y}(y')\equiv\de_{yy'}$, which is $1$ if $y'=y$ and $0$ otherwise. This expresses $q$ as a convex sum of mutually orthogonal measures since $\de_{y}\perp\de_{y'}$ for all $y\ne y'$. Set 
\be
N_{q}:=\{y\in Y\;:\;q_{y}=0\}
\ee 
and let $Y\xstoch{\psi}X$ be a disintegration of $(\phi,p,q)$. 
Then $p$ also decomposes as 
\be
\label{eq:pdecomposesviadisint}
p=\sum_{y\in Y}q_{y}\psi_{y}\equiv\sum_{y\in Y\setminus N_{q}}q_{y}\psi_{y},
\ee
where the set of probability measures $\{\psi_{y}\}_{y\in Y\setminus N_{q}}$ are mutually orthogonal 
because $\psi_{y}$ is a measure supported on $f^{-1}(\{y\})$. 
Furthermore, $\phi$ preserves the mutual orthogonality of these measures
\be
\label{eq:phipreservesorthog}
(\phi\circ \psi_{y})\perp(\phi\circ \psi_{y'})\qquad\forall\;y\ne y'\in Y\setminus N_{q},  
\ee
since $\phi\circ \psi_{y}=\de_{y}$ for all $y\in Y\setminus N_{q}$. 
Setting $\w_{y}$ to be the state corresponding to $\psi_{y}$ gives 
\be
\begin{split}
H_{f}(\omega)&\overset{\text{(\ref{eq:pdecomposesviadisint})}}{=\joinrel=\joinrel=}
H_{f}\left(\sum_{y\in Y\setminus N_{q}}q_{y}\w_{y}\right)
\underset{\text{Defn~\ref{defn:orthogonallyaffinefibredfunctor}}}{\overset{\text{(\ref{eq:phipreservesorthog})}}{=\joinrel=\joinrel=\joinrel=\joinrel=\joinrel=}}
\sum_{y\in Y\setminus N_{q}}q_{y}H_{f}(\w_{y})\\
&\overset{\text{Lem~\ref{lem:Hcoboundary}}}{=\joinrel=\joinrel=\joinrel=\joinrel=}\sum_{y\in Y\setminus N_{q}}q_{y}\Big(H_{\mA}(\w_{y})-H_{\mB}(\underbrace{\w_{y}\circ f}_{\de_{y}})\Big)
=\sum_{y\in Y\setminus N_{q}}q_{y}H_{\mA}(\w_{y})
\ge0.
\end{split}
\ee
The last line holds because $H_{\mB}$ vanishes on pure states and by the assumption that $H_{\mA}$ is always non-negative. 
\eprf

Proposition~\ref{prop:positivitycommutative} shows that our axioms \emph{imply} the (seemingly strong) axiom of non-negativity for entropy difference used by BFL in their functorial characterization of Shannon entropy (Theorem~\ref{thm:BFL}). Combining this fact with Lemma~\ref{lem:BFLimplieszeroonpure} suggests that it is reasonable to replace the BFL axiom of non-negativity for entropy difference by non-negativity of $H_{\mA}$ and equality to zero on pure states. In fact, a corollary of Proposition~\ref{prop:positivitycommutative} and BFL's characterization is an alternative functorial characterization of Shannon entropy that does not explicitly use the non-negativity for entropy difference assumption. However, we still need one more important fact to show that our notion for a functor being orthogonally affine is equivalent to BFL's notion of a functor being externally affine on finite probability spaces (Proposition~\ref{lem:externalaffineorthogonalaffine}). We will then use this towards building the final fact used in our characterization theorem. 

\begin{lem}[Invariance under adjoining zero]{lem:invarianceadjoinzero}
Let $H:\NCFP\to\B\R$ be an orthogonally affine fibred functor for which $H_{\mA}(\w)\ge0$ for all states $\w\in\mS(\mA)$, with equality on all pure states,  for all $C^*$-algebras $\mA$.
Let $\mA$ and $\mB$ be $C^*$-algebras and let $\pi:\mA\oplus\mB\twoheadrightarrow\mA$ be the projection. Then $H_{\pi}(\omega)=0$ for all $\w\in\mS(\mA)$. In particular, 
if $X$ and $Y$ are finite sets and $\iota:X\hookrightarrow X\amalg Y$ is the inclusion with associated $*$-homomorphism $\pi:\C^{X\amalg Y}\twoheadrightarrow\C^{X}$, then $H_{\pi}(\w)=0$ for all states $\w\in\mS(\C^{X})$. 
\end{lem}

\bprf
The proof is similar to that of Lemma~\ref{lem:Hinvariantunderiso} since $\w_{x}\circ\pi$ is pure whenever $\w_{x}$ is. 
\eprf

\begin{prop}[External versus orthogonal affinity]{lem:externalaffineorthogonalaffine}
Let $H:\FinProb^{\op}\to\B\R$ be a fibred functor for which $H_{\mA}(\w)\ge0$ for all states $\w\in\mS(\mA)$, with equality on all pure states, for all commutative $C^*$-algebras $\mA$. Then $H$ is orthogonally affine 
if and only if $H$ is externally affine.
\end{prop}

\bprf

\noindent
($\Rightarrow$)
Suppose $H$ is orthogonally affine. The external convex sum of $(\C^{X'},\w')\xrightarrow{f}(\C^{X},\w)$ and $(\C^{Y'},\xi')\xrightarrow{g}(\C^{Y},\xi)$ defines a morphism 
\be
\left(\C^{X'\amalg Y'},\l\widetilde{\w}'+(1-\l)\widetilde{\xi}'\right)\xrightarrow{k:=f\oplus g}\left(\C^{X\amalg Y},\l\widetilde{\w}+(1-\l)\widetilde{\xi}\right),
\ee
where the tildes denote the states as viewed on the direct sum (cf.\ Example~\ref{ex:BFLexternal}). In particular, $(\C^{X}\oplus\C^{Y},\widetilde{\w})\xrightarrow{\pi_{X}}(\C^{X},\w)$ is a morphism in $\NCFP$ for example. Furthermore, 
\be
\widetilde{\w}\circ k=\widetilde{\w}',\qquad
\widetilde{\xi}\circ k=\widetilde{\xi}',\qquad
\widetilde{\w}\perp\widetilde{\xi},\qquad\text{ and }\qquad
\widetilde{\w}'\perp\widetilde{\xi}', 
\ee
which says that $f\oplus g$ preserves the orthogonality of $\widetilde{\w}$ and $\widetilde{\xi}$. 
Since $H$ is orthogonally affine, 
\be
\begin{split}
H(k)&\equiv H_{f\oplus g}\left(\l\widetilde{\w}+(1-\l)\widetilde{\xi}\right)\\
&\overset{\text{Defn~\ref{defn:orthogonallyaffinefibredfunctor}}}{=\joinrel=\joinrel=\joinrel=\joinrel=\joinrel=}\l H_{f\oplus g}\big(\widetilde{\w}\big)+(1-\l)H_{f\oplus g}\big(\widetilde{\xi}\big)\\
&\overset{\text{Lem~\ref{lem:Hcoboundary}}}{=\joinrel=\joinrel=\joinrel=\joinrel=}\l \Big(H_{\C^{X\amalg Y}}\big(\widetilde{\omega}\big)-H_{\C^{X'\amalg Y'}}\big(\widetilde{\omega}'\big)\Big)
+(1-\l)\Big(H_{\C^{X\amalg Y}}\big(\widetilde{\xi}\big)-H_{\C^{X'\amalg Y'}}\big(\widetilde{\xi}'\big)\Big)\\
&\overset{\text{Lem~\ref{lem:invarianceadjoinzero}}}{=\joinrel=\joinrel=\joinrel=\joinrel=}\l\Big(H_{\C^{X}}(\w)\!-\!H_{\C^{X'}}(\omega')\Big)+(1\!-\!\l)\Big(H_{\C^{Y}}(\xi)\!-\!H_{\C^{Y'}}(\xi')\Big)\\
&\overset{\text{Lem~\ref{lem:Hcoboundary}}}{=\joinrel=\joinrel=\joinrel=\joinrel=}\l H_{f}\left({\w}\right)+(1-\l)H_{g}\left({\xi}\right)
\equiv \l H(f)+(1-\l)H(g).
\end{split}
\ee

\noindent
($\Leftarrow$)
Suppose $H$ is externally affine. 
Let $p$, $q$ be probability measures on $X$ and let $p'$, $q'$ be probability measures on $X'$. Let $X\xrightarrow{\phi}X'$ be a function that preserves both pairs of probability measures, i.e.\ $\phi\circ p=p'$ and $\phi\circ q=q'$. Suppose $p\perp q$ as well as $p'\perp q'$. In what follows, we will first show that there exist morphisms $(A,p_{\restriction A})\xrightarrow{\psi}(A',p'_{\restriction A'})$ and $(B,q_{\restriction B})\xrightarrow{\eta}(B',q'_{\restriction B'})$ such that $\l\psi\oplus(1-\l)\eta=\phi$. 
Let $S_{r}$ denote the support of $r\in\{p,q,p',q'\}$ (viewed as a subset of $X$ or $X'$ depending on the subscript). By assumption, $S_{p}\cap S_{q}=\varnothing$ and $S_{p'}\cap S_{q'}=\varnothing$. Furthermore, $\phi$ can be visualized as 
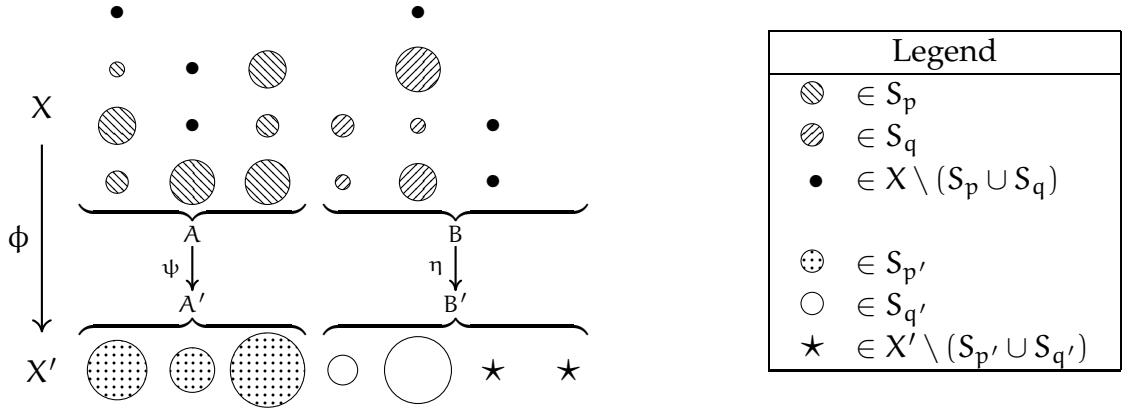
\begin{center}
    \begin{tikzpicture}
    \node at (-1,3) {$X$};
    \node at (-1,-0.5) {$X'$};
        \draw[-{>[scale=2.5,
          length=2,
          width=3]},thick] (-1,2.5) -- node[left]{$\phi$} (-1,0);
    \draw[pattern=dots] (0,-0.5) circle (2.0ex);
    \draw[pattern=dots] (1.0,-0.5) circle (1.5ex);
    \draw[pattern=dots] (2.0,-0.5) circle (2.5ex);
    \draw (3.0,-0.5) circle (1.0ex);
    \draw (4.0,-0.5) circle (2.25ex);
    \node at (5.0,-0.5) {\Large$\star$};
    \node at (6.0,-0.5) {\Large$\star$};
    \draw[pattern=north west lines] (0,2.0) circle (0.75ex);
    \draw[pattern=north west lines] (0,2.75) circle (1.25ex);
    \draw[pattern=north west lines] (0,3.5) circle (0.5ex);
    \node at (0,4.25) {$\bullet$};
    \draw[pattern=north west lines] (1.0,2.0) circle (1.5ex);
    \node at (1.0,2.75) {$\bullet$};
    \node at (1.0,3.5) {$\bullet$};
    \draw[pattern=north west lines] (2.0,2.0) circle (1.5ex);
    \draw[pattern=north west lines] (2.0,2.75) circle (0.75ex);
    \draw[pattern=north west lines] (2.0,3.5) circle (1.25ex);
    \draw[pattern=north east lines] (3.0,2.0) circle (0.5ex);
    \draw[pattern=north east lines] (3.0,2.75) circle (0.75ex);
    \draw[pattern=north east lines] (4.0,2.0) circle (1.25ex);
    \draw[pattern=north east lines] (4.0,2.75) circle (0.5ex);
    \draw[pattern=north east lines] (4.0,3.5) circle (1.5ex);
    \node at (4.0,4.25) {$\bullet$};
    \node at (5.0,2.0) {$\bullet$};
    \node at (5.0,2.75) {$\bullet$};
    \node at (1.0,1.45) {$\underbrace{\hspace{30mm}}_{A}$};
    \node at (1.0,0.25) {$\overbrace{\hspace{30mm}}^{A'}$};
    \node at (4.5,1.45) {$\underbrace{\hspace{35mm}}_{B}$};
    \node at (4.5,0.25) {$\overbrace{\hspace{35mm}}^{B'}$};
    \draw[-{>[scale=2.5,
          length=1.75,
          width=2]},thick] (1.0,1.15) -- node[left]{\scriptsize$\psi$} (1.0,0.55);
    \draw[-{>[scale=2.5,
          length=1.75,
          width=2]},thick] (4.5,1.15) -- node[left]{\scriptsize$\eta$} (4.5,0.55);
    \node at (11.0,1.75) {\begin{tabular}{|c|}\hline Legend\\\hline\begin{tabular}{rl}\begin{tikzpicture}\draw[pattern=north west lines] (0,2.0) circle (0.75ex);\end{tikzpicture}&$\in S_{p}$\\\begin{tikzpicture}\draw[pattern=north east lines] (0,2.0) circle (0.75ex);\end{tikzpicture}&$\in S_{q}$\\$\bullet$&$\in X\setminus(S_{p}\cup S_{q})$\\ & \\\begin{tikzpicture}\draw[pattern=dots] (0,2.0) circle (0.75ex);\end{tikzpicture}&$\in S_{p'}$\\\begin{tikzpicture}\draw (0,2.0) circle (0.75ex);\end{tikzpicture}&$\in S_{q'}$\\\Large$\star$&$\in X'\setminus(S_{p'}\cup S_{q'})$\end{tabular}\\\hline\end{tabular}};
    \end{tikzpicture}
\end{center}
where the indicated sets are defined by 
\be
A':=S_{p'},\qquad B':=S_{q'}\cup\big(X\setminus (S_{p'}\cup S_{q'})\big),\qquad
A:=\phi^{-1}(A'),\qquad B:=\phi^{-1}(B'), 
\ee
and the functions $A\xrightarrow{\psi}A'$ and $B\xrightarrow{\eta}B'$ are defined by restricting $\phi$ to $A$ and $B$, respectively. If we also define the probability measures $p_{\restriction A}, q_{\restriction B}, p'_{\restriction A'},$ and $q'_{\restriction B'}$ on $A, B, A',$ and $B'$, respectively, then $(A,p_{\restriction A})\xrightarrow{\psi}(A',p'_{\restriction A'})$ and $(B,q_{\restriction B})\xrightarrow{\eta}(B',q'_{\restriction B'})$ are morphisms in $\FinProb$ and most importantly, 
\be
\l\left(\!
\xy0;/r.25pc/:
(0,7.5)*+{\big(A,p_{\restriction A}\big)}="1";
(0,-7.5)*+{\big(A',p'_{\restriction A'}\big)}="2";
{\ar"1";"2"^{\psi}};
\endxy
\!\right)\oplus(1-\l)\left(\!
\xy0;/r.25pc/:
(0,7.5)*+{\big(B,q_{\restriction B}\big)}="1";
(0,-7.5)*+{\big(B',q'_{\restriction B'}\big)}="2";
{\ar"1";"2"^{\eta}};
\endxy
\!\right)
=
\xy0;/r.25pc/:
(0,7.5)*+{\big(X,\l p+(1-\l)q\big)}="1";
(0,-7.5)*+{\big(X',\l p'+(1-\l)q'\big)}="2";
{\ar"1";"2"^{\phi}};
\endxy
.
\ee
Thus, 
\be
\begin{split}
H_{\phi}\big(\l p+(1-\l)q\big)&\equiv H\big(\l\psi\oplus(1-\l)\eta\big)\\
&\overset{\text{Defn~\ref{defn:externallyaffinefunctor}}}{=\joinrel=\joinrel=\joinrel=\joinrel=}\l H(\psi)+(1-\l)H(\eta)\\
&=\l\Big(1 H(\psi)+0 H(\eta)\Big)+(1-\l)\Big(0 H(\psi)+1H(\eta)\Big)\\
&\overset{\text{Defn~\ref{defn:externallyaffinefunctor}}}{=\joinrel=\joinrel=\joinrel=\joinrel=}\l H(1\psi\oplus0\eta)+(1-\l)H(0\psi\oplus1\eta)\\
&\equiv\l H_{\phi}(p)+(1-\l)H_{\phi}(q), 
\end{split}
\ee
which completes the proof. 
\eprf

\begin{rmk}[External affinity ignores the internal structure of quantum states]{rmk:externalignoresinternal}
The objects of $\FinProb$ are convex generated by the single object $\mathbf{1}$, which is the (essentially) unique probability space consisting of a single element.   Indeed, an arbitrary finite probability space $(X,p)$ can be decomposed into an external convex sum as
$
(X,p)\cong\bigoplus_{x\in X}p_{x}\mathbf{1}.
$
However, in $\NCFP$, a non-commutative probability space such as $(\mathcal{M}_{m},\omega)$ cannot be expressed as an external convex combination of lower-dimensional probability spaces. 
Therefore, the statement ``if $H$ is externally affine (on all $C^*$-algebras), then $H$ is orthogonally affine'' is false.%
\footnote{Although the converse is still true, as can be seen by a minor modification of the proof of the $(\Rightarrow)$ direction in Proposition~\ref{lem:externalaffineorthogonalaffine}.}
Example~\ref{ex:externallyaffine}~(\ref{item:quantumShannondifference}) is a counter-example because it is not orthogonally affine. 
This, together with Proposition~\ref{lem:externalaffineorthogonalaffine} provides some motivation for our choice of defining convex structures \emph{internally} on the fibres over $C^*$-algebras. 
\end{rmk}

\begin{cor}[Characterizing the Shannon entropy on commutative $C^*$-algebras]{cor:commutativecharacterization}
Suppose $H:\NCFP\to\B\R$ is a continuous orthogonally affine fibred functor for which $H_{\mA}(\w)\ge0$ for all states $\w\in\mS(\mA)$, with equality on all pure states, for all $C^*$-algebras $\mA$. Then there exists a constant $c\ge0$ such that $H_{f}=cS_{f}$ for all $*$-homomorphisms $f$ between \emph{commutative} $C^*$-algebras. 
\end{cor}

\bprf
Continuity and functoriality are already assumed. Non-negativity of $H_{f}(\w)$ for all states $\w$ and $*$-homomorphisms between commutative $C^*$-algebras was proved in Proposition~\ref{prop:positivitycommutative}. Finally, the notion of affine orthogonality of $H$ is equivalent to external affinity for commutative $C^*$-algebras by Proposition~\ref{lem:externalaffineorthogonalaffine}. By BFL's characterization theorem (Theorem~\ref{thm:BFL}), $H$ is the functor giving the difference of entropies on the subcategory of commutative probability spaces up to an overall non-negative constant. 
\eprf

The orthogonally affine assumption for \emph{all} $C^*$-algebras will provide the last fact needed to prove our characterization theorem. 

\begin{lem}[Affine orthogonality determines entropy]{lem:affineorthogtoclassical}
Let $H:\NCFP\to\B\R$ be a continuous and orthogonally affine fibred functor for which $H_{\mA}(\w)\ge0$ for all states $\w\in\mS(\mA)$, with equality on all pure states,  for all $C^*$-algebras $\mA$.
If $\w$ is any state on $\mA$, 
then there exists a constant $c\ge0$ (independent of the algebras and states) such that 
$H_{\mA}(\w)=cS(\w)$.
\end{lem}

\bprf
By invariance of $H$ under $*$-isomorphisms (Lemma~\ref{lem:Hinvariantunderiso}), it suffices to assume $\w$ is a state 
as in Example~\ref{ex:expectationvalues}. 
Let $N_{p}:=\{x\in X\,:\, p_{x}=0\}$ be the nullspace of $p$. For each $x\in X\setminus N_{p}$, decompose $\w_{x}$ into a convex sum $\w_{x}=\sum_{y\in Y_{x}}\psi_{yx}\w_{yx}$ of pure states $\w_{yx}\in\mS(\mM_{m_{x}})$, where $\{\psi_{yx}\}_{y\in Y_{x}}$ defines a nowhere-vanishing probability measure on a finite set $Y_{x}$ whose cardinality equals the rank 
of the support of $\w_{x}$. Thus, $X\setminus N_{p}\xstoch{\psi}\coprod_{x\in X\setminus N_{p}}Y_{x}$ defines a stochastic map. Let $P_{yx}\in\mM_{m_{x}}$ denote the one-dimensional projection associated to the pure state $\w_{yx}$. If $P_{x}$ denotes the support of $\w_{x}$, then $P_{x}=\sum_{y\in Y_{x}}P_{yx}$ for all $x\in X\setminus N_{p}$. Set
$\mB:=\left(\bigoplus_{x\in X\setminus N_{p}}\C^{Y_{x}}\right)\oplus\C^{\{\bullet\}}$, 
where $\C^{\{\bullet\}}\cong\C$, and $\bullet$ merely serves as a label to distinguish it from the rest of the algebra. 
Define a $*$-homomorphism $\mB\xrightarrow{f}\mA$ by 
\be
\C^{Y_{x}}\ni e_{y}\xmapsto{f} \left(\bigoplus_{x'\in X\setminus\{x\}}0\right)\oplus P_{yx}
\quad\text{ and }\quad
\C^{\{\bullet\}}\ni e_{\bullet}\xmapsto{f}\left(\bigoplus_{x\in X\setminus N_{p}}(\mathds{1}_{m_{x}}-P_{x})\right)\oplus\bigoplus_{x\in N_{p}}\mathds{1}_{m_{x}}, 
\ee
where the first case expresses $P_{yx}$ as an element of $\mB$ (with $0$'s on all factors other than $\mM_{m_{x}}$). Then $f$ is a (unital) $*$-homomorphism that preserves the orthogonality of \emph{all} the $\w_{yx}$ states with $y\in Y_{x}$ and $x\in X\setminus N_{p}$ (by viewing all the $\omega_{yx}$ as states on $\mA$ via Lemma~\ref{lem:invarianceadjoinzero}). Therefore, 
\be
\label{eq:Hdisint}
\begin{split}
H_{\mA}(\w)-H_{\mB}(\w\circ f)
&=H_{f}(\w)
=\sum_{x\in X\setminus N}\sum_{y\in Y_{x}}p_{x}\psi_{yx} H_{f}(\w_{yx})\\
&=\sum_{x\in X\setminus N}\sum_{y\in Y_{x}}p_{x}\psi_{yx}\big(H_{\mA}(\w_{yx})-H_{\mB}(\w_{yx}\circ f)\big)
=0
\end{split}
\ee
because $\w_{yx}$ and $\w_{yx}\circ f$ are pure states. 

Consequently, 
\be
\begin{split}
H_{\mA}(\w)&\overset{\text{(\ref{eq:Hdisint})}}{=\joinrel=\joinrel=}H_{\mB}(\w\circ f)\\
&{\overset{\text{Cor~\ref{cor:commutativecharacterization}}}{=\joinrel=\joinrel=\joinrel=\joinrel=}}-c\sum_{x\in X\setminus N_{p}}\sum_{y\in Y_{x}}p_{x}\psi_{yx}\log(p_{x}\psi_{yx})\quad\text{ for some $c\ge0$}\\
&=-c\!\!\!\!\sum_{x\in X\setminus N_{p}}\underbrace{\sum_{y\in Y_{x}}\psi_{yx}}_{1}p_{x}\log(p_{x})-c\!\!\!\!\sum_{x\in X\setminus N_{p}}\!\!\!p_{x}\!\sum_{y\in Y_{x}}\psi_{yx}\log(\psi_{yx})\\
&=c\left(S(p)+\sum_{x\in X}p_{x}S(\w_{x})\right), 
\end{split}
\ee
where the last equality follows from the definition of the Shannon entropy for the $S(p)$ term and Lemma~\ref{lem:derivationpropertyonorthogonal} for the $S(\omega_{x})$ term. 
\eprf

\begin{theo}[A functorial characterization of quantum entropy]{thm:abstractentropy}
Let $H:\NCFP\to\B\R$ be a continuous and orthogonally affine fibred functor
for which $H_{\mA}(\w)\ge0$ for all states $\w\in\mS(\mA)$, with equality on all pure states, for all $C^*$-algebras $\mA$.
Then there exists a constant $c\ge0$ such that 
\[
H_{f}(\omega)=c\Big(S(\omega)-S(\omega\circ f)\Big)
\]
for all morphisms $\mB\xrightarrow{f}\mA$ of $C^*$-algebras and states $\omega\in\mS(\mA)$. 
\end{theo}

\bprf
Since $H_{f}(\w)=H_{\mA}(\w)-H_{\mB}(\w\circ f)$ by Lemma~\ref{lem:Hcoboundary}, Lemmas~\ref{lem:Hinvariantunderiso} and~\ref{lem:affineorthogtoclassical} show this equals the entropy difference up to the same constant $c$. 
\eprf

It is interesting that the notion of a disintegration was used in the proof of Proposition~\ref{prop:positivitycommutative}. Note that in the category of states on (finite-dimensional) $C^*$-algebras and state-preserving $*$-homomorphisms, disintegrations do not always exist~\cite{PaRu19}. Nevertheless, when they exist, they imply $H_{f}(\omega)\ge0$, as the following proposition shows. Since the definition of a non-commutative disintegration is not needed anywhere else in this work, the reader is referred to~\cite{PaRu19} for definitions and other facts assumed in the proof. 

\begin{prop}[If a disintegration for $(f,\w,\omega\circ f)$ exists, then $S_{f}(\w)\ge0$]{prop:disintimpliesnonneg}
Let $\mB\xrightarrow{f}\mA$ be a $*$-homomorphism and $\mA\xstoch{\w}\C$ a state on $\mA$. If $(f,\w,\omega\circ f)$ has a disintegration, then $S_{f}(\w)\ge0$. 
\end{prop}

\bprf
By isomorphism invariance of $S$, it suffices to consider the case where $\mA,\mB,\w,f$, and $\xi$ are as in Lemma~\ref{lem:ptracedirectsums} (without the unitaries $U_{x}$). Let $N_{p}\subset X$ and $N_{q}\subset Y$ be the nullspaces of $p$ and $q$, respectively. Assume that a disintegration of $(f,\omega,\xi)$ exists. By the non-commutative disintegration theorem~\cite[Theorem~5.76]{PaRu19}, for each $x\in X$ and $y\in Y$ there exist non-negative matrices $\t_{yx}\in\mathcal{M}_{c_{xy}}$ such that 
\be
\label{eq:taujimatrices}
\tr\left(\sum_{x\in X}\t_{yx}\right)=1\qquad\forall\;y\in Y\setminus{N}_{q}
\quad\text{ and }\quad
p_{x}\rho_{x}=\bigboxplus_{y\in Y}\t_{yx}\otimes q_{y}\s_{y}
\qquad\forall\;x\in X. 
\ee
One more fact that will be needed is the equality
\be
\label{eq:logoftensor}
(C\otimes D)\log(C\otimes D)=C\log(C)\otimes D+C\otimes D\log(D)
\ee
for all non-negative square matrices $C,D$ (possibly of different sizes). 
Computing $S_{\mA}(\omega)$ first gives 
\be
\begin{split}
S_{\mA}(\omega)&\overset{\text{Defn~\ref{defn:entropyquantum}}}{=\joinrel=\joinrel=\joinrel=\joinrel=\joinrel=}-\sum_{x\in X}\tr\big(p_{x}\rho_{x}\log(p_{x}\rho_{x})\big)\\
&\overset{\text{(\ref{eq:taujimatrices})}}{=\joinrel=\joinrel=\joinrel=}-\sum_{x\in X}\tr\!\left(\bigboxplus_{y\in Y\setminus N_{q}}\!\!(\tau_{yx}\otimes q_{y}\s_{y})\log\!\left(\bigboxplus_{y'\in Y\setminus N_{q}}\!\!\!\tau_{y'x}\otimes q_{y'}\s_{y'}\!\right)\!\!\right)\\
&=-\sum_{x\in X}\sum_{y\in Y\setminus N_{q}}\tr\big((\tau_{yx}\otimes q_{y}\s_{y})\log(\tau_{yx}\otimes q_{y}\s_{y})\big)\\
&\overset{\text{(\ref{eq:logoftensor})}}{=\joinrel=\joinrel=\joinrel=}-\!\sum_{x\in X}\sum_{y\in Y\setminus N_{q}}\!\!\!\tr\big(\t_{yx}\log(\t_{yx})\otimes q_{y}\s_{y}+\t_{yx}\otimes q_{y}\s_{y}\log(q_{y}\s_{y})\big)\\
&\overset{\text{(\ref{eq:taujimatrices})}}{=\joinrel=\joinrel=\joinrel=}
\sum_{y\in Y\setminus N_{q}}q_{y}S\left(\bigboxplus_{x\in X}\t_{yx}\right)+S_{\mB}(\xi), 
\end{split}
\ee
where $\bigboxplus_{x\in X}\t_{yx}$ is viewed as a density matrix on $\mM_{s_{x}}$, where $s_{x}:=\sum_{y\in Y\setminus N_{q}}c_{yx}$.  
Thus, 
\be
S_{f}(\w)=S_{\mA}(\w)-S_{\mB}(\xi)=\sum_{y\in Y\setminus N_{q}}q_{y}S\left(\bigboxplus_{x\in X}\t_{yx}\right)\ge0. \qedhere
\ee
\eprf

\begin{rmk}[Having a disintegration is not necessary for $S_{f}(\w)\ge0$]{rmk:disintDNimplySfpos}
If $S_{f}(\w)\ge0$, it is not necessary that a disintegration of $(f,\w,\omega\circ f)$ exists. A counter-example is the inclusion $\mM_{2}\to\mM_{2}\otimes\mM_{2}$, which sends $B\in\mM_{2}$ to $\mathds{1}_{2}\otimes B$, and the density matrix 
$
\rho=
\mathrm{diag}(p_{1},p_{2},p_{3},p_{4}),
$
where $p_{1},p_{2},p_{3},p_{4}\ge0$ satisfy 
$p_{1}+p_{2}+p_{3}+p_{4}=1,$ 
$p_{1}+p_{3}>0,$ and $p_{2}+p_{4}>0.$
Then 
\[
S_{f}(\w)=-p_{1}\log\left(\frac{p_{1}}{p_{1}+p_{3}}\right)
-p_{2}\log\left(\frac{p_{2}}{p_{2}+p_{4}}\right)
-p_{3}\log\left(\frac{p_{3}}{p_{1}+p_{3}}\right)
-p_{4}\log\left(\frac{p_{4}}{p_{2}+p_{4}}\right)
\ge0,
\]
while a disintegration exists if and only if  $p_{1}p_{4}=p_{2}p_{3}$~\cite[Example~4.25]{PaRu19}. 
\end{rmk}


\begin{rmk}[A brief history and comparison of axiomatizations of quantum entropy]{rmk:historyentropy}
Quantum entropy and its variants were often built upon the classical versions, whose many axiomatizations are reviewed in Csiszar's survey~\cite{Cs08}. In 1932, von~Neumann obtained a phenomelogical characterization of entropy~\cite[Chapter~V.\ Section~2]{vN18}. 
In 1968, Ingarden and Kossakowski characterized the von~Neumann entropy using dimensional partial Boolean rings of projections in Hilbert space~\cite{InKo68}. In 1974, Ochs provided a characterization using partial isometric invariance, additivity, subadditivity, and continuity (plus some additional technical axioms)~\cite{Oc75}. In 1975, Thirring~\cite{Th75} characterized the von~Neumann entropy using axioms closely related to those implemented by Fadeev in his characterization of the Shannon entropy~\cites{Fa56,Fa57}, the latter of which was simplified by Renyi~\cite{Re61}.%
\footnote{Thirring's statement and proof can be found in~\cite[(2.2.4) pages 58--61]{Th83}. However, it seems that the first written account of his proof in English appears in Wehrl's review~\cite[pages 238--239]{We78}.}

Thirring's characterization is most closely related to ours and it is worth taking the time to spell out his assumptions, which read as follows. 
\vspace{-2mm}
\begin{enumerate}[(i)]
\itemsep0pt
\item
$S(\rho)$ is a continuous function of the eigenvalues of $\rho$; 
\item
$S(\frac{1}{2}\mathds{1}_{2})=\log 2$; 
\item
If $\Hi=\bigoplus_{n=1}^{N}\Hi_{n}$ is a direct sum of Hilbert spaces and if $\rho=\bigoplus_{n=1}^{N}p_{n}\rho_{n}$ is a weighted direct sum of density matrices, where $\{p_{n}\}_{n\in\{1,\dots,N\}}$ is a probability distribution on $\{1,\dots,N\}$, then $S(\rho)=S(p)+\sum_{n=1}^{N}p_{n}S(\rho_{n}),$ where $p$ is viewed as a diagonal matrix on $\C^{N}$ with entries given by the $p_{n}$. 
\end{enumerate}
\vspace{-2mm}
There are actually several implicitly hidden assumptions within these three. For example, the dependence on eigenvalues means $S(\rho)=S(U\rho U^{\dag})$ for all unitaries $U$, i.e.\ $S(\rho)$ is invariant under $*$-isomorphisms. The second item is merely a normalization condition, which we have ignored (it specifies the constant $c$). The third item is close to our orthogonally affine assumption. However, an implicit assumption is made, which can be expressed as saying that $S(\rho_{n})$ is equal to $S(0\oplus\cdots\oplus\rho_{n}\oplus\cdots0)$, i.e.\ $S$ is invariant under the non-unital inclusion of one matrix algebra into a direct sum. This is closely related to Och's partial isometry invariance assumption. In our characterization, we obtain this property \emph{as well as} invariance under $*$-isomorphisms as a consequence of our axioms. 

Two other characterizations of the von~Neumann entropy have appeared recently. The first is the topos-theoretic one of Constantin and D\"oring, which is based on how different commutative subalgebras, called \emph{contexts}, of a fixed $C^*$-algebra determine its structure~\cite{CoDo20}. A context may be thought of as probing a quantum system by measurements of an observable and sending any state to the probability measure on the associated set of eigenvalues---in other words, it is a $*$-homomorphism. The collection of all contexts forms a category via inclusion and one can define measures associated to this category via compatible families of probability measures on the contexts \emph{without} defining a state on the embedding algebra. They then classify the quantum entropy by assuming the form of entropy on the subcategory of commutative algebras and minimizing over all contexts. One difference between our assumptions for characterizing the von~Neumann entropy is that we do not assume the formula for the Shannon entropy, nor do we assume that commutative algebras play any special role, which are singled out by the existence of disintegrations for \emph{all} state-preserving *-homomorphisms. On the other hand, their characterization emphasizes the physically intuitive operational importance of classical systems for determining the entropy. 

Finally, there has also been an abstract characterization of the von~Neumann entropy by homological information structures introduced by Baudot and Bennequin (cf.\ Theorem~3 page 3290 and Theorem~4 page 3313 of~\cite{BaBe15}), which are further developed by Vigneaux~\cites{Vi19,Vi20}. They seek to understand information quantities more generally. It is not yet clear to us how our methods are related. 
\end{rmk}

\subsubsection*{Acknowledgements}
The work presented in this manuscript began while the author was at the CUNY Graduate Center, continued when the author was at the University of Connecticut, and was completed at the Institut des Hautes \'Etudes Scientifiques. 
First, the author thanks Tobias Fritz, who
explained several aspects of his work
with John Baez and Tom Leinster~\cite{BFL}
and who provided insight and additional references.
The author has benefited from conversations with 
Jonathan Ben-Benjamin, 
Lewis Bowen,
Tai-Danae Bradley, 
Brian Dressner, 
James Fullwood, 
Brian Hall, 
Azeem ul~Hassan, 
Chris Heunen, 
Mark Hillery, 
Manas Kulkarni, 
Franklin Lee, 
Jamie Lennox, 
William Mayer, 
Vadim Oganesyan, 
Philip Parzygnat, 
George Poppe, 
Xing Su, 
Josiah Sugarman, 
Dennis Sullivan,
Steven Vayl, 
Scott O. Wilson, 
Cody Youmans, 
and Lai-Sang Young. 
The author thanks Anders Kock and two anonymous referees for helpful feedback on an earlier version of this work. 
The author thanks Yung Bae for motivation. 
Finally, and most importantly, the author is especially thankful to V.~P.~Nair,
who provided several crucial suggestions during the earlier stages of this work. 
This work was partially supported by 
NSF grant PHY-1213380 
and
the Capelloni Dissertation Fellowship.
This research has also received funding from the European Research Council (ERC) under the European Union's Horizon 2020 research and innovation program (QUASIFT grant agreement 677368).

\addcontentsline{toc}{section}{\numberline{}Bibliography}
\bibliography{vNentropyBib}

\Addresses

\end{document}